\documentstyle[12pt,psfig]{article}
\renewcommand{\theequation}{\thesection.\arabic{equation}}
\newlength{\extraspace}
\newlength{\extraspaces}
\newcommand{\be}{\begin{equation}
\addtolength{\abovedisplayskip}{\extraspaces}
\addtolength{\belowdisplayskip}{\extraspaces}
\addtolength{\abovedisplayshortskip}{\extraspace}
\addtolength{\belowdisplayshortskip}{\extraspace}}
\newcommand{\ee}{\end{equation}}
 
\newcommand{\ba}{\begin{eqnarray}
\addtolength{\abovedisplayskip}{\extraspaces}
\addtolength{\belowdisplayskip}{\extraspaces}
\addtolength{\abovedisplayshortskip}{\extraspace}
\addtolength{\belowdisplayshortskip}{\extraspace}}
\newcommand{\ea}{\end{eqnarray}}
 
\newcommand{\bas}{\begin{eqnarray*}
\addtolength{\abovedisplayskip}{\extraspaces}
\addtolength{\belowdisplayskip}{\extraspaces}
\addtolength{\abovedisplayshortskip}{\extraspace}
\addtolength{\belowdisplayshortskip}{\extraspace}}
\newcommand{\eas}{\end{eqnarray*}}
 
\newcounter{subequation}[equation]
\makeatletter
\expandafter\let\expandafter\reset@font\csname reset@font\endcsname
\def\subeqnarray{\arraycolsep1pt
    \def\@eqnnum\stepcounter##1{\stepcounter{subequation}
        {\reset@font\rm(\theequation\alph{subequation})}}\eqnarray}

\makeatother
 
\newenvironment{theorem}[1]
{\vspace{3mm}\noindent {\bf #1 :} }{\vspace{2mm}}
\newcommand{\bt}[1]{\begin{theorem}{#1}}
\newcommand{\et}{\end{theorem}}
 
\newcommand{\newsection}[1]{
\vspace{12mm}
\pagebreak[3]
\addtocounter{section}{1}
\setcounter{equation}{0}
\setcounter{subsection}{0}
 
\begin{flushleft}
{\large\bf \thesection. #1}
\end{flushleft}
\nopagebreak
\medskip
\nopagebreak}
 
\newcommand{\newsubsection}[1]{
\vspace{1cm}
\pagebreak[3]
 
\addtocounter{subsection}{1}
\noindent{ \bf \thesubsection. #1}
\nopagebreak
\vspace{2mm}
\nopagebreak}
 
\newcommand{\NP}[1]{Nucl.\ Phys.\ {\bf #1}}

\newcommand{\CMP}[1]{Comm.\ Math.\ Phys.\ {\bf #1}}
\newcommand{\PR}[1]{Phys.\ Rev.\ {\bf #1}}

\begin{document}
%
\begin{titlepage}
%
\renewcommand{\thefootnote}{\fnsymbol{footnote}}
\begin{flushright}
BN-TH-96-10\\
NTZ 33/96\\
August 1996
\end{flushright}
\vspace{1cm}
 
\begin{center}
{\Large {\bf Gauge parameter dependence and gauge invariance}} \\[4mm]
{\Large {\bf in the Abelian Higgs model}}
{\makebox[1cm]{  }       \\[1.5cm]
{\bf Rainer H\"au\ss ling }\\ [3mm]
{\small\sl Institut f\"ur Theoretische Physik und} \\
{\small\sl Naturwissenschaftlich-Theoretisches 
           Zentrum, Universit\"at Leipzig} \\
{\small\sl Augustusplatz 10/11, D-04109 Leipzig, Germany} \\[0.5cm]
{and}\\[0.5cm]
{\bf Elisabeth Kraus} \footnote{Supported by 
Deutsche Forschungsgemeinschaft}\\ [3mm]
{\small\sl Physikalisches Institut, Universit\"at Bonn} \\
{\small\sl Nu\ss allee 12, D-53115 Bonn, Germany}} 
\vspace{1.5cm}
 
{\bf Abstract}
\end{center}
\begin{quote}
We analyze gauge parameter dependence 
by using an algebraic method which relates
the gauge parameter dependence of Green functions
to an enlarged Slavnov-Taylor identity. In the course of the renormalization
it turns out that gauge parameter dependence of physical parameters is 
already restricted at the level of Green functions. In a first step
we consider the on-shell conditions which we find to be in
complete agreement with these restrictions to all orders of perturbation
theory. The fixing of the coupling, however, is much more involved outside
the complete on-shell scheme. In the Abelian Higgs model we prove that
this fixing can be properly chosen by 
requiring the Ward identity of gauge invariance
to hold in its tree form to all orders of perturbation theory.
\end{quote}
\vfill
\renewcommand{\thefootnote}{\arabic{footnote}}
\setcounter{footnote}{0}
\end{titlepage}
%
\newsection{Introduction}
In the perturbative formulation of gauge theories one has to destroy
gauge symmetry by fixing the gauge. Thereby arbitrary gauge parameters
are introduced into the action which have to be proven to cancel when 
physical observables are concerned. The Green functions, however, depend
non-trivially on these gauge parameters. Already in the early days of
non-abelian gauge theories it has been suggested that the S-matrix is
a gauge parameter independent quantity \cite{lee}. After having
established the renormalization of gauge theories via the BRS symmetry
gauge parameter independence of the S-matrix has been strictly proven
in \cite{BRS2} for such gauge theories which are completely broken by
the spontaneous symmetry breaking mechanism and which therefore do not
contain any massless particles. The ingredients of this proof are the complete
on-shell scheme fixing the poles and the residua of the particles at
the physical masses and a specific on-shell normalization condition for
the transversal vector-vector-Higgs vertex. In order to prove gauge
parameter independence of the S-matrix within this special set of
normalization conditions heavy technical tools as the Wilson operator
product expansion are required. In pure  gauge
theories with massless gauge bosons the situation is much more
involved because the S-matrix cannot be simply
constructed due to infrared divergencies. For these theories it has been
derived among other things that the $\beta$-functions are independent
of the gauge parameter. In this context, the original proof explicitly 
refers to an invariant scheme, namely the minimal
subtraction scheme of dimensional regularization \cite{KSZ}. 

Aiming at the standard model such considerations are
unsatisfactory: Due to the masslessness of the photon a complete
on-shell scheme, especially for the $W$-propagator, cannot be formulated
in a strict sense, and an invariant scheme is not available due to the
parity violating interactions in the fermion sector. Furthermore, in concrete
calculations one is not only interested in the gauge parameter
independence of the S-matrix but also in a knowledge, how the single
Green functions depend on the gauge parameters.  We therefore analyze 
gauge parameter dependence by an algebraic method
\cite{PS}, which can be generally applied in the context of Green functions.
Although we eventually aim at the standard model we present the techniques
for the Abelian Higgs model here as a natural first step.

The algebraic method
enlarges the usual BRS transformations by varying the gauge
parameter into a Grassmann variable. Constructing the Green functions in
agreement with the enlarged BRS invariance automatically yields
information about gauge parameter dependence. Moreover, it is seen that the
physical parameters of the model 
have to be chosen gauge-parameter-independently
in order to be able to prove gauge parameter independence of
the S-matrix. For higher orders in
perturbation theory these restrictions are extended to restrictions for the
normalization conditions and it is seen that the conditions of
ref.~\cite{BRS2} for spontaneously broken theories  and the
MS-scheme for symmetric theories \cite{KSZ} are just a special set of adequate
normalization conditions (cf.~sec.~5 and ref.~\cite{PS}, \cite{KS2}, where the
respective
analysis has been performed for pure gauge theories).

The paper is organized as follows: In section 2 we present the
Abelian Higgs model and also summarize the results of
ref.~\cite{KS}.  In section 3 the enlarged BRS transformations and
Slavnov-Taylor identities are introduced, which give rise to the
algebraic control of gauge parameter dependence. In section 4 the
classical approximation is solved, thereby finding the restrictions for 
the physical parameters mentioned above. In section 5 these
considerations are continued to higher orders. This is an
application of the method and shows in an impressive way, how the
algebraic method works in higher orders. Furthermore, one deduces 
in this context 
that the on-shell conditions are in complete agreement with the enlarged
Slavnov-Taylor identity. In section 6 and 7 Ward identites of rigid and
local symmetry are derived including the BRS-varying gauge parameter.
As a consequence of the local Ward identity it turns out that gauge
parameter dependence of the longitudinal part of the 3-point vertex,
which describes the interaction between vector, Higgs and would-be
Goldstone, is competely determined by the gauge dependence of the self
energies of the scalars. Section 8 contains a summary of the results.
In appendix~A we give the general classical solution of the Abelian
Higgs model including all the free parameters. As an illustration, in
appendix~B we sketch the diagrams contributing to the gauge
parameter variation of the Higgs self energy.

\newsection{The Abelian Higgs model}
In order to set the general framework and to fix notation we first give a 
short exposition of the model treated in this paper.\\
The Abelian Higgs model contains a vector field $A_{\mu}$ and two
scalar fields $\underline \varphi = (\varphi_1, \varphi_2)$
 interacting in such a way that
$U(1)$ gauge invariance is broken spontaneously. It can be described
by the following classical action (given in conventional normalization):
\begin{equation}
\label{gl1.1}
\Gamma_{inv} = \int \left\{ -\frac{1}{4} F_{\mu \nu} F^{\mu \nu} +
    \frac{1}{2} (D_{\mu} \underline \varphi ) (D^{\mu} \underline \varphi) -
    \frac{1}{8} \frac{m_H^2}{m^2} e^2 
      (\varphi_1^2 + 2\frac me \varphi_1 +
\varphi _2^2 )^2 \right\}
\end{equation}
with
\begin{equation}
\label{gl1.2}
F_{\mu \nu} \equiv \partial_{\mu} A_{\nu} - \partial_{\nu} A_{\mu} \; , \;
 D_{\mu} \varphi_1 \equiv \partial_{\mu} \varphi_1 + eA_{\mu} \varphi_2
\; , \; 
D_{\mu} \varphi_2 \equiv \partial_{\mu} \varphi_2 - eA_{\mu} (\varphi_1 + 
\frac me)
\end{equation}
$\Gamma_{inv}$ is invariant under the $U(1)$ transformations
\begin{equation}
\label{gl1.3}
  \delta_{\omega} A_{\mu} = \partial_{\mu} \omega
\; , \;
\delta_{\omega} \varphi_1 = -e\omega \varphi_2 \; , \;
\delta_{\omega} \varphi_2 = e\omega (\varphi_1 + \frac me )
 \; .
\end{equation}
 The shift  $\frac me$ of the field $\varphi_1 $ produces the mass $m$ for  
 the vector field $A_{\mu}$  
 and $\varphi_1$ is the (physical) Higgs field with mass
$m_H$. The field $\varphi_2$ then takes the role of the would-be Goldstone
boson eaten up by $A_{\mu}$.\\[0.5cm]
In order to quantize the model the gauge has to be fixed,
\begin{equation}
\label{gl1.4}
\Gamma_{g.f.} = \int \left\{ \frac{1}{2} \xi B^2 +
  B (\partial A + \xi_A m \varphi_2) \right\} \; . 
\end{equation}
$\xi$ denotes the gauge parameter and $B$ is an auxiliary field with
$\delta_{\omega} B = 0$. 
Please note that the 't Hooft gauge fixing term $\int B \xi_A m \varphi_2$
has been introduced in order to avoid a non-integrable infrared 
singularity in the $<\! \varphi_2 \varphi_2 \! >$ 
propagator. It leads to a non-trivial
violation of both local and global gauge invariance:
\begin{equation}
\label{gl1.5}
\delta_{\omega} \Gamma_{g.f.} = \int \left\{ \omega \Box B +
  e \omega B \xi_A m (\varphi_1 + \frac me ) \right\}
\end{equation}
Hence one has to enlarge local gauge transformations to BRS
transformations thereby introducing the Faddeev-Popov ($\phi \pi$) fields
$c, \bar{c}$,
\begin{eqnarray}
\label{gl1.6}
s A_{\mu} = \partial_{\mu} c & , & sc = 0 \; \; , \nonumber \\
s \varphi _1 = - e c \varphi_2 & , & 
s \varphi _2 =  e c \bigl(\varphi_1 +\frac me\bigr)
 \; \; , \\
s  \bar{c} = B & , & s B = 0 \; \; , \nonumber 
\end{eqnarray}
and to postulate BRS invariance of the theory instead of gauge invariance:
\begin{equation}
\label{gammabrs}
s \Gamma_{cl} = 0
\end{equation}
This leads to extra-terms in the classical action:
\begin{equation}
\label{gl1.7}
{\Gamma}_{cl} = \Gamma_{inv} + \Gamma_{g.f.} + \Gamma_{\phi \pi}
  + \Gamma_{e.f.}
\end{equation}
The $\phi \pi$-part 
\begin{equation}
\label{gl1.8}
\Gamma_{\phi \pi} = \int \left\{ - \bar{c} \Box c -
  e \bar{c} \xi_A m (\varphi_1 + \frac{m}{e} ) c \right\}
\end{equation}
is chosen such that its BRS variation exactly cancels the BRS variation of
$\Gamma_{g.f.}$.\\
The external field part
\begin{equation}
\label{gl1.9}
\Gamma_{e.f.} = \int \left\{ Y_1 (s\varphi_1) + Y_2 (s\varphi_2) \right\}
\end{equation}
couples the non-linear BRS transformations $s \varphi_i $ to the
external fields $Y_i$.
It is added in order to allow for a careful definition 
of the BRS transformation of $\varphi_i$ in higher orders (cf.~(\ref{gl1.10})).
 BRS invariance has been shown to define the 
Abelian Higgs model in its quantized version and it is the relevant
symmetry for renormalizability and unitarity of the S-matrix 
\cite{BRS}.\\[0.5cm]
In a foregoing paper \cite{KS} rigid and also local gauge symmetry have been
formulated to all orders of perturbation theory (see sections 6 and 7 for
details).  
For this purpose the gauge fixing has to be complemented by  external
scalar fields $\hat \varphi_1$ and $\hat \varphi_2$ which transform 
under gauge transformations according to
\begin{equation}
\label{efg}
\delta _{\omega} \hat \varphi_1 = - e \omega \hat \varphi_2 \; , \;
\delta _{\omega} \hat \varphi_2 = e \omega (\hat \varphi_1 -\xi _A \frac me)
\; .
\end{equation}
Then the enlarged gauge fixing  
\begin{equation}
\label{ggf}
\Gamma_{g.f.} = \int \left\{ \frac{1}{2} \xi B^2 +
  B \partial A - e 
B\Bigl( (\hat \varphi_1 -  \xi_A \frac me )\varphi_2  -
\hat \varphi_2 ( \varphi_1 -  \hat \xi_A \frac me ) \Bigr) \right\}  
\end{equation}  
is invariant under the global gauge transformations (\ref{gl1.3}) and 
(\ref{efg}), if one adjusts
$\hat \xi _A  =  -1  $. For vanishing external fields 
$\hat{\varphi}_i$ the original
gauge fixing (\ref{gl1.4}) is recovered and these fields have only
been introduced in order to be able to manage the breaking of
gauge invariance  for higher orders Green functions algebraically.\\
Also in its enlarged form gauge invariance
does not describe the $\phi \pi $-part, even it does not in the
classical approximation.
Instead, BRS invariance has to be used for the construction of
the model, which involves also the external fields $\hat \varphi _i $.
Because they are coupled to BRS variations they are transformed
into  external fields $q_i$ with $\phi\pi $-charge $+1$:
\begin{equation}
s\hat \varphi_i  = q_i \; , \; s q_i = 0 \; , \; i = 1,2
\end{equation}
At the level of the vertex functional ${\Gamma} = {\Gamma}_{cl} +
O(\hbar)$ the BRS invariance of the theory is encoded in the 
Slavnov-Taylor (ST) identity
\begin{equation}
\label{gl1.10}
{\cal S}( {\Gamma}) \equiv \int \left\{ 
  \partial_{\mu} c \frac{\delta  {\Gamma}}{\delta A_{\mu}} +
  B \frac{\delta  {\Gamma}}{\delta \bar{c}} +
\frac{\delta  {\Gamma}}{\delta \underline{Y}}
  \frac{\delta  {\Gamma}}{\delta \underline{\varphi}}+
\underline{q} \frac{\delta  {\Gamma}}{\delta \hat{\underline{\varphi}}}
\right\} = 0
\end{equation}
where $\underline F = (F_1 ,F_2 ) $.
The ST identity is the essential ingredient for the proof of renormalizability
and unitarity.   
It can be proven that (\ref{gl1.10}) together with 
appropriate normalization conditions, invariance under charge conjugation 
and the gauge condition (\ref{ggf}) uniquely defines the model  to all orders
of perturbation theory. (The quantum numbers of all fields are given in table
1.)
\begin{table}
\begin{center}
\begin{tabular}{l|c|c|c|c|c|c|c|c|c|c}
fields & $A_{\mu}$ & $B$ & $\tilde{\varphi}_1$ & $\tilde{\varphi}_2$ &
  $c$ & $\bar{c}$ & $Y_1$ & $Y_2$ & $q_1$ & $q_2$ \\ \hline
dim    & 1 & 2 & 1 & 1 & 0 & 2 & 3 & 3 & 1 & 1 \\ \hline
charge conj. & - & - & + & - & - & - & + & - & + & - \\ \hline 
$Q_{\phi \pi}$ & 0 & 0 & 0 & 0 & +1 & -1 & -1 & -1 & +1 & +1 
\end{tabular}
\end{center}
{{\sl Table 1}: Quantum numbers of the fields ($\tilde{\varphi}_i =
\varphi_i, \hat{\varphi}_i$)}
\end{table}  
In the first step one has to show that the classical action (\ref{gl1.7})
is uniquely determined as the local solution of the ST identity. The 
general classical solution has been calculated in ref.\  \cite{KS} and is
presented in appendix A explicitly. The free para\-meters appearing in the
general solution are the usual field and coupling renormalizations given
by:
\begin{eqnarray}
\label{fieldred}
& \varphi _i \longrightarrow  \sqrt{ z_i} (\varphi_i  -x _i \hat \varphi_i ) 
\; \;  , \; \;  A_\mu \longrightarrow \sqrt{z_A} A_\mu & \\
& m \longrightarrow \sqrt {z_m} m \; \; , \; \;   
  m_H \longrightarrow \sqrt {z_{m_H} } m_H \; \; , \; \;
  e \longrightarrow z_e e & \nonumber
\end{eqnarray}
The field redefinitions of the remaining fields and parameters are 
governed by the ST identity and the gauge fixing (\ref{ggf}).
Via the field redefinitions the external fields $\hat \varphi_i $ enter the
gauge invariant part $\Gamma _{inv}$ (\ref{gl1.1}) of the action, but
only in the combination
\begin{equation}
\label{barphi}
\bar \varphi_i = \varphi _i - x_i \hat \varphi_i \; ,
\end{equation}
with $ x_i$, $i = 1,2$, being further free parameters.
These parameters appear also in the BRS variations of the fields
$\varphi_i $, which modify (\ref{gl1.6}) into
\begin{equation}
\label{brsmod}
s \varphi_1 =-e\bar \varphi _2 c + x_1 q_1 \; \; , \; \;
   s \varphi_2 = e (\bar \varphi _1 + \frac me)c
+ x_2 q_2 \; .
\end{equation}
The free parameters  have to be fixed by appropriate normalization
conditions order by order in perturbation theory.
In order to have a proper field theoretic definition of  the S-matrix
the masses of the particles have to be fixed at the poles of the
2-point Green functions; these conditions read for the 
corresponding vertex functions:
\begin{eqnarray}
\label{massnorm}
Re \; \Gamma_{\varphi_1 \varphi_1} (p^2 = m_H^2 ) = 0 &\mbox{fixes} &  z_{m_H} 
 \nonumber \\
\Gamma^T (p^2 = m^2) = 0 & \mbox{fixes} &   z_m   \\
\Gamma_{c \bar{c}} (p^2 = m_{\hbox{ghost}}^2) = 0 & \mbox{fixes} &
  \xi_A \nonumber
\end{eqnarray}
Thereby we have defined the transversal part of the  2-point function
according to
\begin{equation}
\label{transvec}
 \Gamma _{A^\mu A^\nu} (p,-p) \equiv \Gamma_{\mu \nu} (p,-p) 
= (\eta_{\mu \nu} -
  \frac{p_{\mu} p_{\nu}}{p^2}) \Gamma^T (p^2) + \frac{p_{\mu} p_{\nu}}{p^2}
  \Gamma^L(p^2) \; .
\end{equation}
The wave function renormalizations $z_i $ are determined on the residua of the
respective 2-point functions,
\begin{eqnarray}
\label{wavenorm}
\partial_{p^2} \Gamma^T (p^2 = m^2) = 1 & \mbox{fixes} & z_A \nonumber \\
Re \; \partial_{p^2} \Gamma_{\varphi_1 \varphi_1} (p^2 = m_H^2) = 1
  & \mbox{fixes} & z_1  \\
\partial_{p^2} \Gamma_{\varphi_2 \varphi_2} (p^2 = \kappa^2) = 1
  & \mbox{fixes} & z_2 \; , \nonumber 
\end{eqnarray}
whereas the parameters $x_i $ are chosen to be fixed on the external field
part  at an arbitrary normalization point $\kappa $:
\begin{eqnarray}
\label{xnorm}
\Gamma_{Y_1 q_1} (p^2 = \kappa^2) = x_1^{(0)} & \mbox{fixes} & x_1  \\
\Gamma_{Y_2 q_2} (p^2 = \kappa^2) = x_2^{(0)} & \mbox{fixes} & x_2 \nonumber 
\end{eqnarray}
$x_1^{(0)}$ and $x_2^{(0)}$ are assumed to be pure numbers (e.g. 1 or 0)
throughout the paper. Using rigid invariance they are not  independent
but $x_2^{(0)} =x_1^{(0)} + O(\hbar) $ (cf.~(\ref{xdef})).\\
It remains to give an appropriate normalization condition for the coupling 
$e$. This coupling can be determined at the 3-point function $\Gamma_{A_{\mu}\varphi_1
\varphi_2 } $  at a normalization momentum $p_{norm}$:
\begin{equation}
\label{coupnorm}
\partial_{p_1^\nu} \Gamma_{A_\mu \varphi_1 \varphi_2 } 
(-p_1-p_2,p_1,p_2) \Big|_{\{p_i\} = p_{norm} } =  - i e \eta^{\mu \nu}
  \quad \mbox{fixes} \quad z_e
\end{equation}
Finally we have to require the 2-dimensional BRS-invariant scalar 
field polynomial (see appendix A) to vanish:
\begin{equation}
\label{vacnorm}
\Gamma _{\varphi_1}
 = 0 \quad\hbox{fixes} \quad \mu
\end{equation}
Applying these normalization conditions to the tree approximation
and requiring the gauge condition (\ref{ggf}) 
 the original classical action $\Gamma_{cl}$ (\ref{gl1.7}) is recovered
up to the modifications encoutered in (\ref{barphi}) and (\ref{brsmod}),
i.e.~$z_a = 1+ \delta
z_a$ with $\delta z_a$ of order $\hbar$.\\[0.5cm]
When controlling the gauge parameter algebraically (the subject of this paper) one gets 
restrictions on the gauge parameter dependence
of the free parameters $z_a$. In higher orders these restrictions
have to be extended to a proper definition of the normalization conditions.
The main result of this paper 
is the observation that the on shell conditions (\ref{massnorm}) for
the physical particles are
in complete agreement with the restrictions given on gauge parameter
dependence, whereas the definition of the coupling (\ref{coupnorm})
has to be modified 
in an appropriate way. It turns out that the  Ward identity
of local symmetry  (cf.~(\ref{wardloc})) can
be used to fix  the gauge parameter dependence of the  vertex 
$ \Gamma_{A_\mu \varphi_1 \varphi_2 } $ correctly.

\newsection{Algebraic control of gauge parameter dependence}
Let us now turn to the proper subject of this paper: the control of 
gauge parameter dependence. 
We start with the observations that for the classical action 
${\Gamma}_{cl}$ the gauge parameter dependence is given by 
a BRS variation\footnote{Eq. (\ref{gl2.1}) is the deeper 
reason for introducing the auxiliary field $B$.}
\begin{equation}
\label{gl2.1}
\partial_{\xi} {\Gamma}_{cl} = \frac{1}{2} \int B^2 =
  \frac{1}{2} s \int \bar{c} B
\end{equation}
and that the r.h.s. therefore vanishes 
between physical states, i.e. physical quantities as the S-matrix are
gauge parameter independent in the tree approximation as they should.
Now we can ask whether 
it is possible to extend this statement to higher  orders. There the differentiation with respect to $\xi$ produces a non-trivial insertion, 
\begin{equation}
\label{binsert}
\partial_{\xi} \Gamma = \Bigl[\int \frac{1}{2} B^2 + O(\hbar)
\Bigr]_4 \cdot \Gamma \; ,
\end{equation}
which cannot be handled in such a simple way as it was 
the case in the tree approximation.
(Note that the vertex $B^2$ can be inserted non-trivially into loop diagrams
according to the appearance of the mixed $B$-$A_\mu$ propagator.)
To prove $\xi $-independence of physical quantities demands quite a technical
effort  \cite{BRS2}, if one does not use specifically adapted tools.
Such a tool has been provided by ref.~\cite{PS} and consists in
BRS-transforming $\xi$ into a Grassmann variable $\chi$ with 
 $\phi \pi$-charge $+1$:
\begin{equation}
\label{gl2.2}
s \xi = \chi \; \; , \; \; s \chi = 0
\end{equation}
Accordingly the ST identity (\ref{gl1.10}) has to be modified into:
\begin{equation}
\label{gl2.3}
{\cal S} (\Gamma) + \chi \partial_{\xi} \Gamma = 0
\end{equation}
Differentiating this last eq. with respect to $\chi$ and evaluating at
$\chi = 0$ leads to
\begin{equation}
\label{gl2.4}
-s^{\chi = 0}_{\Gamma}  \partial _\chi \Gamma  \Big|_{\chi = 0}
+ \left. \partial_{\xi} \Gamma \right|_{\chi = 0} = 0 \; .
\end{equation}
In the model under investigation $s_{\Gamma}$ is defined in (\ref{gl4.15})
and is
 -- roughly speaking -- the functional generalization of
$s$. Hence (\ref{gl2.4}) is nothing else but the functional analog of
(\ref{gl2.1}), which can be -- in contrast to (\ref{binsert}) -- controlled
in higher orders. In other words: Proving (\ref{gl2.3})
to all orders of perturbation theory automatically yields 
the $\xi$-dependence of the 1-PI Green functions in an algebraic way.
 
In theories with spontaneous breaking of the symmetry (as in the 
Abelian Higgs model) even a second gauge parameter $\xi_A$ has to be
introduced via the 't Hooft term.
For the purpose of this paper we
restrict ourselves to the consideration of the dependence on the gauge parameter $\xi$ only and do not BRS-transform $\xi _A $. However, we allow that 
$\xi_A $ might be a function of $\xi$. For this reason
 the 't Hooft gauge, which is defined by the choice
\begin{equation}
\xi_A = \xi \; ,
\end{equation}
is automatically included in our discussion.

We want to finish this section by writing down how the 
classical action (\ref{gl1.7}) has to be modified due to 
the BRS-transforming gauge parameter $\xi $ (\ref{gl2.2}).
The only place, where $\xi$ appears,  is the gauge fixing term (\ref{ggf}) 
and -- in the
't Hooft gauge -- the $\phi\pi$-part.
Evaluating the BRS variations we end up with
\begin{equation}
\Gamma_{g.f.} + \Gamma_{\phi\pi} \longrightarrow \Gamma_{g.f.} 
+ \Gamma_{\phi\pi}+ \chi \int
\left\{ \frac 12 \bar c B + (\partial_{\xi} \xi _A) m \bar c \varphi_2 -
(\partial_{\xi} \hat \xi_A) m \bar c\hat \varphi _2  \right\} \; .
\end{equation}
Note that $\chi \bar c B $ as well as $\chi \bar c \varphi_2 $ can be  
non-trivially
inserted into loop diagrams. Therefore the
renormalization of the model for $\chi \neq 0 $ has to be carried
out carefully to all orders
looking thereby into the modifications appearing from gauge parameter
dependence. 

\newsection{Slavnov-Taylor identity for $\chi \neq 0$}
Following  the remarks of the last section we are able to control
the gauge parameter dependence of the Green functions, if we
construct them
with the help of the
 $\chi$-enlarged ST identity\footnote{From here on we use the
symbol ${\cal S}$ for all the differential operators on the l.h.s. of
(\ref{gl2.3}).}
\begin{equation}
\label{gl3.2}
{\cal S} (\Gamma ) \equiv \int \left\{ 
  \partial_{\mu} c \frac{\delta \Gamma}{\delta A_{\mu}} + 
  B \frac{\delta \Gamma}{\delta \bar{c}} + 
  \frac{\delta \Gamma}{\delta \underline{Y}} 
   \frac{\delta \Gamma}{\delta \underline{\varphi}} + 
  \underline{q} \frac{\delta \Gamma}{\delta \underline{\hat{\varphi}}}
  \right\} + \chi \partial_{\xi} \Gamma = 0 \; .
\end{equation}
Because the
ST identity does not prescribe the gauge fixing terms, we can 
also postulate the gauge condition (\ref{ggf}),
\begin{equation}
\label{gl3.3}
\left. \frac{\delta \Gamma}{\delta B} \right|_{\chi = 0} = \xi B \; + \;
  \partial A \; - \;
  e \left[ (\hat{\varphi}_1 - \xi_A \frac{m}{e}) \varphi_2 -
           \hat{\varphi}_2 (\varphi_1 - \hat{\xi}_A \frac{m}{e} ) 
	   \right] \; ,
\end{equation}
to hold for the solution $\Gamma$ of (\ref{gl3.2}).  
It is worth to note that
(\ref{gl3.3}) is well defined and can be integrated in all orders of 
perturbation theory due to the linearity in the propagating fields.

Because $\chi $ is a Grassmann variable and therefore $\chi ^2  = 0 
$, the vertex functional $\Gamma$ can be decomposed into a 
$\chi$-independent and an explicitly $\chi$-dependent part,
\begin{equation}
\label{gl3.4}
\Gamma = \hat{\Gamma} + \chi Q \; ,
\end{equation}
where  $Q$ is the generating functional of Green functions with
$\phi\pi$-charge $-1$.
Inserting this ansatz into (\ref{gl3.2}), making use of
$\chi^2 = 0$, and looking for terms proportional to $\chi^0$ and $\chi^1$
one immediately gets:
\begin{eqnarray}
\label{gl3.5a}
& \chi^0 : {\displaystyle \int} \left\{ 
  \partial_{\mu} c \displaystyle \frac{\delta \hat{\Gamma}}{\delta A_{\mu}}
+ B \displaystyle \frac{\delta \hat{\Gamma}}{\delta \bar{c}}
+ \displaystyle \frac{\delta \hat{\Gamma}}{\delta \underline{Y}} 
     \displaystyle \frac{\delta \hat{\Gamma}}{\delta \underline{\varphi}}
+ \underline{q} 
     \displaystyle \frac{\delta \hat{\Gamma}}{\delta \underline{\hat{\varphi}}}
\right\} = 0 & \\
\label{gl3.5b}
& \! \! \! \chi^1 : {\displaystyle \int} \left\{
  \partial_{\mu} c \displaystyle \frac{\delta (\chi Q)}{\delta A_{\mu}}
+ B \displaystyle \frac{\delta (\chi Q)}{\delta \bar{c}}
+ \displaystyle \frac{\delta (\chi Q)}{\delta \underline{Y}} 
    \displaystyle \frac{\delta \hat{\Gamma}}{\delta \underline{\varphi}}
+ \displaystyle \frac{\delta \hat{\Gamma}}{\delta \underline{Y}} 
    \displaystyle \frac{\delta (\chi Q)}{\delta \underline{\varphi}}
+ \underline{q} 
    \displaystyle \frac{\delta (\chi Q)}{\delta \underline{\hat{\varphi}}} 
\right\} + \chi \partial_{\xi} \hat{\Gamma} = 0 & 
\end{eqnarray}
Equation (\ref{gl3.5a}) is exactly the ST identity (\ref{gl3.2}) for
$\chi = 0$, which has been studied in \cite{KS}.
Therefore we only have to consider (\ref{gl3.5b})
and to investigate the modifications brought about by the BRS-varying
gauge parameter via the $\chi$-dependent part $Q$.

First we have to look for the 
general classical solution of the ST identity
(\ref{gl3.5b}) in order to find the free parameters of the
model.
The general solution $\hat{\Gamma}^{gen}_{cl}$ has already been presented
in section 2 and is given explicitly  in appendix A.
In the
classical approximation
the decomposition (\ref{gl3.4}) implies that   $Q$ is a local field polynomial 
with dimension less than or
equal to four, carries $\phi \pi$-charge $-1$  and is even under
charge conjugation. The most general ansatz for $Q_{cl}$ is therefore given
by (see table of quantum numbers)
\begin{eqnarray}
\label{gl3.6}
Q_{cl} & = & {\textstyle \int} \{ d_1 Y_1 \varphi_1 \; + \;
                     \hat{d}_1 Y_1 \hat{\varphi}_1 \; + \;
                     d Y_1 \; + \;
d_2 Y_2 \varphi_2 \; + \; \hat{d}_2 Y_2 \hat{\varphi}_2 \nonumber \\
& & \; \; + \; f \bar{c} \varphi_2 \; + \;
                \hat{f} \bar{c} \hat{\varphi}_2 \; + \;
\tilde{f} \bar{c} B \; + \; f_A \bar{c} \partial A \nonumber \\
& & \; \; + \; h_1 \bar{c} \varphi_1 \varphi_2 \; + \;
                h_2 \bar{c} \hat{\varphi}_1 \varphi_2 \; + \;
h_3 \bar{c} \varphi_1 \hat{\varphi}_2 \; + \;
h_4 \bar{c} \hat{\varphi}_1 \hat{\varphi}_2 \} \; ,
\end{eqnarray}
and the 13 parameters $d_1, \hat{d}_1, d ,\dots ,h_3 ,h_4$ have to be 
determined with the help of (\ref{gl3.5b}). This calculation is 
straightforward and most easily done by differentiating (\ref{gl3.5b})
with respect to suitable fields and then comparing coefficients of
independent terms. We find:
\begin{equation}
\label{gl3.7}
Q_{cl} = Q_{e.f.} + Q_{\phi\pi} 
\end{equation}
with ($x_1^{(0)} = x_2^{(0)} \equiv x$ classically, see (\ref{xdef}),
and $\bar \varphi_i = \varphi_i - x \hat \varphi _i$)
\begin{eqnarray}
\label{gl3.8a}
Q_{e.f.} & = & \int \left\{ \frac{1}{4} (\partial_{\xi} \mbox{ln} z_1
  + \partial_{\xi} \mbox{ln} z_2 ) (Y_1 \bar{\varphi}_1 +
                                    Y_2 \bar{\varphi}_2) \right. \\
& &  \; \; \; \; \left. + \frac{1}{4} (\partial_{\xi} \mbox{ln} z_1
  - \partial_{\xi} \mbox{ln} z_2 ) (Y_1 \bar{\varphi}_1 -
                                    Y_2 \bar{\varphi}_2) 
  - \partial_{\xi} x (Y_1 \hat{\varphi}_1 + Y_2 \hat{\varphi}_2 )                                                             
      \right\} \; , \nonumber \\
\label{gl3.8b}
Q_{\phi\pi} & = & \int \left\{ -\frac{1}{4} e \bar{c}
                (\partial_{\xi} \mbox{ln} z_1
  + \partial_{\xi} \mbox{ln} z_2 ) 
  \left( (\bar{\varphi}_1 + \frac{\sqrt{z_m}}{\sqrt{z_1} z_e}
                            \frac{m}{e} ) \hat{\varphi}_2 -
          \bar{\varphi}_2 (\hat{\varphi}_1 - \xi_A \frac{m}{e}) \right)
  \right. \\
& & \; \; \; \; \left.  - \frac{1}{4} e \bar{c}
                (\partial_{\xi} \mbox{ln} z_1
  - \partial_{\xi} \mbox{ln} z_2 )
  \left( (\bar{\varphi}_1 + \frac{\sqrt{z_m}}{\sqrt{z_1} z_e}
                            \frac{m}{e} ) \hat{\varphi}_2 +
         \bar{\varphi}_2 (\hat{\varphi}_1 - \xi_A \frac{m}{e}) \right)
  \right. \nonumber \\
& & \; \; \; \;  + (\partial_{\xi} \xi_A ) m \bar{c} \bar{\varphi}_2
   + \frac 12 \bar c B  \biggr\} \;  \nonumber 
\end{eqnarray}

The parameters $z_1, z_2, z_m, z_e, x$ as well as $\xi_A$ 
in (\ref{gl3.8a}), (\ref{gl3.8b}) are the free
parameters appearing in the 
general solution $\hat{\Gamma}$ of
the ST identity (\ref{gl3.2}) for $\chi = 0$.
  Please note that the coefficients in $Q_{cl}$
are fully determined as functions of these parameters.
The ST identity (\ref{gl3.5b}) does not only determine $Q_{cl}$, but
even  restricts  the $\xi$-dependence of the  free
parameters: Whereas the wave function renormalizations $z_1, z_2$ and
$x$ are allowed to depend arbitrarily on the gauge parameter,
the parameters $z_e, z_A, z_m, z_{m_H}$ and $\mu^2$ have to be independent of
the gauge parameter $\xi $:
\begin{eqnarray}
\label{gl3.10}
& \partial_{\xi} z_e = 0 \; , \; \partial_{\xi} z_A = 0 \; , & \\
& \partial_{\xi} z_m = 0 \; , \; \partial_{\xi} z_{m_H} = 0 \; , \;
  \partial_{\xi} \mu^2 = 0 \nonumber
\end{eqnarray}
This means that only the $\xi$-independent part of these parameters
 has to  be fixed by normalization conditions, 
the $\xi$-dependent part, however,
 is completely determined by the $\chi$-enlarged
ST identity and one cannot dispose on it by a normalization condition.

Applying the normalization conditions to the general classical
solution of the ST identity  the constraints (\ref{gl3.10})
are trivially fulfilled, because $e , m_H $ and $m$ 
are $\xi$-independent
physical parameters of the model.  If one would take these parameters
 depending on 
$\xi$,
then, of course,
 the S-matrix would depend on the gauge parameter $\xi$.
Such a choice
  is  completely unphysical but  it would be allowed just looking
into the renormalization of the model via the ordinary ST identity.
In the tree approximation it is obvious
that nobody will introduce such a strange $\xi$-dependence into the theory.
In higher orders, however, the couplings have to be fixed at 
some value of non-local Green
functions, and there it is much less obvious and transparent, how one
has to deal with the splitting into $\xi$-dependent and $\xi$-independent
quantities in the normalization conditions.
 A wrong
adjustment of $\xi$-dependent counterterms will introduce $\xi$-dependence 
into the S-matrix in the same way as we have illustrated above for the tree
approximation. The next section is devoted to the analysis, how 
the classical restrictions are continued to higher order perturbation
theory.

\newsection{Gauge parameter dependence of Green functions}
Having completed the characterization of the classical model via the
ST identity the next step would be 
 the proof of the $\chi$-dependent ST identity to all
orders. We don't want to go into the details here, but only mention that, 
according to the considerations in \cite{PS}, the proof of (\ref{gl3.2}) 
(for $\chi \neq 0$) can be reduced to the
proof of the ST identity for $\chi = 0$.  It
 turns out that no anomaly occurs, and once suitable
counterterms are adjusted the $\chi$-enlarged ST identity
(\ref{gl3.2})  is valid in all
orders,
\begin{equation}
\label{st}
{\cal S} (\Gamma) = 0 \; ,
\end{equation}
with $\Gamma $ being the generating functional of 1-PI Green functions, which
 includes also the $\chi$--dependent contributions.
  Accordingly the validity of (\ref{st}) will
be assumed throughout the following.

In order to extend the restrictions on gauge parameter dependence found
in the classical approximation (\ref{gl3.10}) to higher orders
we have to consider the
$\xi$-dependence of the  non-local
 Green functions, on which the normalization conditions are applied.
Thereby we want to restrict ourselves to the case, where the Higgs
particle is a stable particle, i.e.~$m_H^2 < 4m^2 $. The imaginary
part is then vanishing for the self energy of the Higgs at $p^2 = m_H^2$
as well as for the transversal part of the vector self energy at 
$p^2 = m^2$ \cite{CLARK}.\\[0.5cm]
$\xi$-independence of the parameter $\mu$ is trivially continued to
higher orders: The ST identity reads for the $\xi$-derivative of 
$\Gamma_{\varphi_1}$ in momentum space:
\begin{equation}
\label{vacnormh}
\partial _\chi \Gamma_{Y_1} (0)  \Gamma_{\varphi_1 \varphi_1}(0) +
\partial_\chi  \Gamma_{Y_1 \varphi_1} (0)  \Gamma_{\varphi_1 } (0) =
- \partial_\xi \Gamma_{\varphi_1}(0)
\end{equation}
Applying the normalization condition (\ref{vacnorm}), which requires
vanishing of the vacuum expectation value of the Higgs field, we end
up with the condition
\begin{equation}
\label{d0}
(- m_H^2 + O(\hbar))  \partial _\chi \Gamma_{Y_1} = 0  \quad \Longrightarrow
\quad \partial _\chi \Gamma_{Y_1} = 0 \; ,
\end{equation}
which is valid to all orders of perturbation theory. \\[0.5cm]
We now turn to the transversal part of the vector 2-point function, 
which was found to be gauge parameter independent in
the classical approximation ($\partial _\xi z_A = 0 $ and $\partial _\xi z_m
 = 0$). This restriction is valid also in higher orders of perturbation
theory: Using the result (\ref{d0}) one gets from the ST identity:
\begin{equation}
\label{vector}
\partial _\chi \Gamma_{Y_2 A_\mu} (p,-p) \Gamma_{\varphi_2 A_\nu} (p,-p)+
\partial _\chi \Gamma_{Y_2 A_\nu} (p,-p) \Gamma_{\varphi_2 A_\mu} (p,-p)=
-\partial_\xi \Gamma_{\mu \nu} (p,-p)
\end{equation}
Because of Lorentz invariance one has
\begin{eqnarray}
\label{lorentz}
\partial _\chi \Gamma_{Y_2 A_\mu} (p,-p)& =&  \frac {p_\mu}m f_1
(\frac {p^2}{m^2}) \; , \\
\Gamma_{\varphi_2 A_\nu} (p,-p) &=& p_\nu m f_2 (\frac {p^2}{m^2}) 
=ip_\nu m + O(\hbar)\; , \nonumber
\end{eqnarray}
and the l.h.s.\ of (\ref{vector}) only contributes to the longitudinal
part of $\partial_\xi \Gamma_{\mu \nu} (p,-p) $. 
Acting with the transversal projector on (\ref{vector}) it is easily seen that 
the transversal part of $\Gamma_{\mu \nu}$ is independent of $\xi$
to all orders of perturbation theory:
\begin{equation}
\label{trans}
\partial _\xi \Gamma^T (p^2) = 0 \quad \hbox{with}\quad
\Gamma_{A_\mu A_\nu} \equiv \Gamma_{\mu \nu} = (\eta_{\mu \nu} -
  \frac{p_{\mu} p_{\nu}}{p^2}) \Gamma^T + \frac{p_{\mu} p_{\nu}}{p^2}
  \Gamma^L  
\end{equation}
This equation has two consequences: It tells us that the non-local
contributions to $\Gamma ^T (p^2) $ are $\xi$-independent,
and also that the counterterms have to
be adjusted in such a way, that (\ref{trans}) is fulfilled 
order by 
order in perturbation theory. Accordingly we have to check whether the
normalization conditions that we have imposed on the transversal part of 
the vector 2-point function (\ref{massnorm}),
(\ref{wavenorm}) are compatible with (\ref{trans}).
Testing (\ref{trans}) at  the normalization points
it is seen that the on-shell condition is in agreement with 
(\ref{trans}):
\begin{eqnarray}
\partial _{p^2} \partial _\xi \Gamma^T (p^2) \Big|_{p^2 = \kappa_m^2}  = \;
\partial _\xi 1 & =& 0 \; , \\
\partial _\xi \Gamma^T (p^2)\Big|_{p^2 = m^2} = \; 
\partial _\xi 0 & =& 0 
\end{eqnarray} 
(Here we have even relaxed  the complete on-shell condition by permitting
an arbitrary normalization point $\kappa_m$ for fixing the residuum.) 
Working in a scheme without explicit normalization conditions like
the MS-scheme (\ref{trans}) has to be proven in concrete calculations
order by order in perturbation theory.
The constraints on the vector 2-point function look very simple and obvious
in the Abelian Higgs model, but in this simple way
 they are  only available 
 by the algebraic control of gauge parameter dependence.\\[0.5cm] 
Much more interesting,
 however, is the 
extension of $\partial _\xi z_{m_H} = 0 $ (\ref{gl3.10}) to higher orders.
 First we consider the wave function renormalization of the scalar particles,
i.e.\ the gauge para\-meter  dependence of $\partial _{p^2}   
\Gamma_{\varphi_i \varphi_i}  (p^2)  $. Testing the ST identity with
respect to the Higgs and the would-be Goldstone we get (using again
(\ref{d0})):
\begin{eqnarray}
\label{higgs}
\partial _\chi \Gamma_{Y_1\varphi_1} (p^2)  \Gamma_{\varphi_1 \varphi_1}(p^2)
&=& - \partial_\xi \Gamma_{\varphi_1 \varphi_1}(p^2) \; , \\
\label{gold}
\partial _\chi \Gamma_{Y_2\varphi_2} (p^2)  \Gamma_{\varphi_2 \varphi_2}(p^2)
&=& - \partial_\xi \Gamma_{\varphi_2 \varphi_2}(p^2) \nonumber
\end{eqnarray}
The normalization conditions on the residua 
(\ref{wavenorm}) uniquely fix the Green functions
$\partial _\chi \Gamma_{Y_i \varphi_i} (p^2)$, explicitly
\begin{equation}
\label{wavefix}
\partial _\chi \Gamma_{Y_i\varphi_i} (\kappa_i^2)  =
-\bigl(
\partial_{p^2} \partial _\chi \Gamma_{Y_i\varphi_i}\bigr)\Big|_{p^2=
\kappa_i^2}  
\Gamma_{\varphi_i \varphi_i}(\kappa_i^2)
\end{equation}
where we have also used a relaxed form of the complete on-shell condition of the
Higgs 2-point function. This equation is well-defined order by order 
in perturbation theory  because $\partial_{p^2}
 \partial _\chi \Gamma_{Y_i\varphi_i} (p^2)  $ is non-local and uniquely
determined from lower orders.
In the complete on-shell scheme (\ref{wavefix}) simplifies 
to\footnote{Therefrom it is immediately
deduced, that the normalization conditions of ref.~\cite{BRS2},
which in addition 
 fix  the coupling 
at an on-shell momentum on the transversal part of $\Gamma_{A_\mu  
A_\nu \varphi_1} $, are 
 in complete agreement with the BRS-varying gauge parameter.}
\begin{equation}
\partial _\chi \Gamma_{Y_i \varphi_i} (m_H^2)  = 0
\end{equation}
Eq.~(\ref{higgs}) completely governs the 
 $\xi$-dependence of the Higgs self-energy.
 However, $\Gamma _{\varphi_1 \varphi_1}
(p^2) $ involves
two free parameters, namely the wave function normalization and the Higgs mass.
Once $
\partial _\chi \Gamma_{Y_1 \varphi_1} (p^2) $ is fixed 
one therefore has to adjust the
mass counterterm in such a way that 
 eq.~(\ref{higgs}) is identically fulfilled.
In contrast to the transversal part of the vector 2-point function the
l.h.s. of (\ref{higgs})
 is far from being  trivial, because 
 the vertex function $
\partial _\chi \Gamma_{Y_1\varphi_1} (p^2)  $ 
 is non-local, i.e. it depends on the momentum
$p^2$, due  to non-trivial insertions
of the vertex $\bar c B $ (see appendix B for the 1-loop diagrams).
These non-local contributions
 govern the $\xi$-dependence of the non-local contributions
to the Higgs self energy.
In 1-loop order eq. (\ref{higgs}) reads
\begin{equation}
\label{higgs1l}
\partial _\chi \Gamma^{(1)}_{Y_1\varphi_1} (p^2) ( p^2 -m_H^2)
= - \partial_\xi \Gamma^{(1)}_{\varphi_1 \varphi_1}(p^2) \; . 
\end{equation}
Testing at $p^2 =m_H^2 $ and using the on-shell condition for the Higgs
(\ref{massnorm}) we see that the l.h.s. as well as the r.h.s. vanish.
This result does not depend on the normalization 
one has chosen for the residuum.\\
Therefore the on-shell conditions are shown to satisfy the
restrictions dictated on gauge parameter dependence via a BRS-varying
gauge parameter. Finally, these arguments can immediately be applied to all
orders of perturbation theory.
This result is quite remarkable and underlines once
more the relevance of the on-shell conditions for the proper field
theoretic definition of spontaneously broken theories.\\[0.5cm] 
We now come to the last constraint: the gauge parameter independence  of
$z_e$ and its compatibility with the normalization condition (\ref{coupnorm}).
It is quite instructive to write down explicitly the equation which
governs the $\xi$-dependence of $\Gamma_{A_\mu \varphi_1 \varphi_2}$:
\begin{eqnarray}
\label{vertex}
\partial_{\chi} \Gamma_{Y_1\varphi _1 } (p_1^2)
\Gamma_{\varphi_1 \varphi_2 A_\mu} (p_1,p_2,p)&+&
\partial_{\chi} \Gamma_{Y_2\varphi _2 } (p_2^2)
\Gamma_{\varphi_1 \varphi_2 A_\mu} (p_1,p_2,p) \nonumber\\
+\partial_{\chi} \Gamma_{Y_2 \varphi _1 A_\mu} (p_2,p_1,p)
\Gamma_{ \varphi_2 \varphi_2} (p_2^2)&+&
\partial_{\chi} \Gamma_{Y_1 \varphi _2 A_\mu} (p_1,p_2,p)
\Gamma_{ \varphi_1 \varphi_1} (p_1^2)  \\
+\partial_{\chi} \Gamma_{Y_2 \varphi _1 \varphi_2} (p,p_1,p_2)
\Gamma_{ \varphi_2 A_\mu} (-p,p) 
& =&  -\partial _\xi 
\Gamma_{\varphi_1 \varphi_2 A_\mu} (p_1,p_2,p) \nonumber 
\end{eqnarray}
This equation completely determines the
$\xi$-dependence of the 3-point vertex. In fact, 
if the residua of the
Higgs and the would-be Goldstone are  fixed
by the normalization conditions (\ref{higgs}) and (\ref{gold}),
eq.~(\ref{vertex}) does not involve any free parameters,
because
all other $\chi$-dependent contributions are non-local. 
It is even not possible to
  evaluate the l.h.s. at a normalization
point as it was the case for the 2-point functions. Therefore, if one insists
in fixing the coupling at a normalization point
as we have done in (\ref{coupnorm}), one is forced to introduce a 
reference point $\xi _o$ in order to fix the gauge parameter independent 
part, and has to govern $\xi$-dependence by eq.~(\ref{vertex}) which
is quite a troublesome task in explicit calculations
(see ref.~\cite{PS} for details).\\
Instead we will show that such a  treatment is not necessary in the
Abelian Higgs model if one agrees to fix the coupling on the local
Ward identity as it is elaborated in section 7.

\newsection{Rigid invariance}
The 
 $\chi$-independent part of the generating functional of 1-PI irreducible
Green functions 
was shown \cite{KS} to satisfy a Ward identity
of rigid symmetry
to all orders of perturbation theory:
\begin{equation}
\label{wgen}
 \hat W^{gen} {\Gamma} \Big|_{\chi =0 } = 0
\end{equation}
with
\begin{eqnarray}
\label{gl4.1}
{\hat W}^{gen}  & \equiv & \int \left\{ 
  - z^{-1} \varphi_2 \frac{\delta}{\delta \varphi_1} +
  z (\varphi_1 - \hat{\xi}_A \frac{m}{e} )
    \frac{\delta}{\delta \varphi_2} -
  z Y_2 \frac{\delta}{\delta Y_1} +
  z^{-1} Y_1 \frac{\delta}{\delta Y_2} \right. \nonumber \\
& & \; \; \; \; \; \left.   - z^{-1} \hat{\varphi}_2
              \frac{\delta}{\delta \hat{\varphi}_1} +
z (\hat{\varphi}_1 - \xi_A \frac{m}{e} )
  \frac{\delta}{\delta \hat{\varphi}_2} -
z^{-1} q_2 \frac{\delta}{\delta q_1} +
z q_1 \frac{\delta}{\delta q_2} \right\}
\end{eqnarray}
The parameters $z,\; \hat \xi_A$ and $\xi_A$ are
 uniquely determined by the
normalization conditions imposed on the residua of the Higgs particle
$\varphi _1$ and the Goldstone boson $\varphi_2 $ (\ref{wavenorm}),
 and the mass normalization
of
the ghosts and the Higgs particle (\ref{massnorm}). They are
expanded in orders of the coupling $e$ according to
the loop expansion,
\begin{equation}
  z = 1 + \delta z
\; , \; \hat \xi _A = -1 + x \xi_A + \delta \hat \xi_A 
\; , \; \xi _A = \xi_A^{(0)}
+ O(\hbar ) \; ,
 \end{equation}
with $ \delta z = O(\hbar) $ and $\delta \hat\xi_A = O(\hbar)  $.
In addition the rigid Ward identity imposes a relation between  the 
vertex function $\Gamma _{Y_1 q_1}$ and $\Gamma _{Y_2 q_2}$. 
Therefore as we have already mentioned, the parameters $x_1$ and $x_2$
are not independent from each other:
\begin{equation}
\label{xdef}
x_2^{(0)} =x_1^{(0)} + O(\hbar) \; \; , \; \;  x\equiv x_1^{(0)} 
\end{equation}

In general the parameters appearing in the Ward identity are functions of
the gauge parameter $\xi$. In fact, in the 't~Hooft gauge $\xi_A$ and 
$\hat \xi _A$ depend on the gauge parameter even in lowest order of
perturbation theory:
\begin{equation}
\label{thooft}
\xi_A = \xi  \; \; , \; \; \hat \xi_A = -1 + x \xi
\end{equation}
We now have the task to study the modifications of (\ref{wgen}), (\ref{gl4.1})
 brought about by a BRS transforming gauge parameter $\xi$.
First we will look at the classical approximation, then to higher 
orders.

\newsubsection{Classical approximation}

The application of the Ward operator $\hat
W^{gen}$ (\ref{gl4.1}) to the general 
classical solution $ \Gamma_{cl}^{gen}$ (\ref{gl3.4}) 
of the ST identity (\ref{gl3.2})
gives a first insight into the modifications expected in higher orders
of perturbation theory:
\begin{equation}
\label{gl4.3}
\hat W^{gen} \Gamma_{cl}^{gen} = \hat
W^{gen} (\hat{\Gamma}_{cl}^{gen} + \chi Q) =
\hat W^{gen} \hat{\Gamma}_{cl}^{gen} + \chi \hat W^{gen} Q = \chi 
\hat W^{gen} Q 
\end{equation}
with ($z = \sqrt{\frac{z_1}{z_2}}$ classically)
\begin{eqnarray}
\label{gl4.4}
\hat W^{gen} Q & = & \int \left\{
  - \partial_{\xi} z (Y_2 + e \bar{c} (\hat{\varphi}_1 -
                                         \xi_A \frac{m}{e}))
                       (\bar{\varphi}_1 
  + \frac{\sqrt{z_m}}{\sqrt{z_1} z_e} \frac me ) \right. \nonumber \\
& & \left. + \; \partial_{\xi} z^{-1} (Y_1 - e \bar{c} \hat{\varphi}_2)
                                   \bar{\varphi}_2 \right. \nonumber \\
& & \left. + \; z \left(
    (\partial_{\xi} (\hat{\xi}_A \frac{m}{e}) -
     x \partial_{\xi} (\xi_A \frac{m}{e}))Y_2 +
    \partial_{\xi} (\xi_A m) \bar{c} (\bar{\varphi}_1 
  + \frac{\sqrt{z_m}}{\sqrt{z_1} z_e} \frac me )
\right) \right\} \neq 0
\end{eqnarray}
The r.h.s. of (\ref{gl4.3}) is  non-vanishing,
because $z$ as well as $\xi _A $ are allowed to depend on the gauge
parameter $\xi $. In the tree approximation  the 't~Hooft gauge
(\ref{thooft}) breaks the naive rigid invariance by
\begin{equation}
\hat
W \Gamma_{cl} = \chi
\int m \bar c  ({\varphi}_1 -x  \hat \varphi _1 + \frac me ) \; .
\end{equation}
This breaking of rigid invariance 
is potentially harmful because of
the appearance of non-linear terms in the propagating (and interacting)
fields, these terms not being well-defined in higher orders and leading
to non-trivial insertions. Therefore one has to absorb these terms
into a functional operator $\chi { V }^{gen} $ which cancels the
unwanted terms when acting on
 $\Gamma_{cl}^{gen}$.
A natural choice is given by extending the operators appearing in
$\hat{W}^{gen}$ as far as possible to invariant operators of the enlarged
$\chi$-varying BRS transformations, i.e.:
\begin{eqnarray}
\label{chiop}
z^{-1} \bigl( \hat \varphi _2 {\delta \over \delta \hat \varphi _1}
 + q _2 {\delta \over \delta  q _1} \bigr)
& \longrightarrow &
z^{-1} \bigl(\hat \varphi _2 {\delta \over \delta \hat \varphi _1} 
 + q _2 {\delta \over \delta  q _1} \bigr)+
\chi \partial _\xi z^{-1} \hat \varphi _2
 {\delta \over \delta q _1} 
\nonumber\\
z \bigl( \hat \varphi _1 {\delta \over \delta \hat \varphi _2}
 + q _1 {\delta \over \delta  q _2} \bigr)
& \longrightarrow &
z \bigl( \hat \varphi _1 {\delta \over \delta \hat \varphi _2} 
 + q _1 {\delta \over \delta  q _2} \bigr)
+ \chi \partial _\xi z \hat  \varphi _1
 {\delta \over \delta q _2} 
\\
z \xi _A \frac me {\delta \over  \delta \hat \varphi_2} & \longrightarrow &
z \xi _A \frac me {\delta \over  \delta \hat \varphi_2}  +
\chi \partial _\xi ( z \xi _A) \frac me {\delta \over  \delta  q_2} \nonumber 
\end{eqnarray}
A short calculation shows that the $\chi$-enlarged Ward operator $W^{gen}$,
\begin{eqnarray}
\label{gl4.12}
W^{gen} & = & \hat W^{gen} + \chi V^{gen} \; , \\
V^{gen} & = & \partial_{\xi} \int \left\{ 
  z (\hat{\varphi}_1 - \xi_A \frac{m}{e}) \frac{\delta}{\delta q_2}
  - z^{-1} \hat{\varphi}_2 \frac{\delta}{\delta q_1} \right\} \; , \nonumber
\end{eqnarray}
removes the harmful breakings leaving only expressions linear
in the propagating fields on the r.h.s., which cannot be inserted non-trivially
into higher orders' loop diagrams:
\begin{equation}
\label{gl4.10}
W^{gen} \Gamma^{gen}_{cl} = \chi \Delta_{br}
\end{equation}
with
 \begin{equation}
\label{gl4.11}
\Delta_{br} = \partial_{\xi} \int \left\{ z^{-1} Y_1 \varphi_2 
                             -z Y_2 (\varphi_1 - \hat{\xi}_A
                             \frac{m}{e}) \right\}
\end{equation}
In the tree approximation when applying the normalization conditions (i.e.\
$z = 1 $)
the 't~Hooft gauge yields:
\begin{equation}
\label{wthooft}
W \Gamma_{cl} = (\hat W - \chi \frac me {\delta \over \delta q_2 }) 
\Gamma_{cl}
= \chi \int Y_2 \frac me 
\end{equation}

\newsubsection{Higher Orders}

Our aim is to demonstrate the validity of the ($\chi$-)deformed WI
to all orders, 
when $W^{gen}$ (\ref{gl4.12}) acts on the generating 
functional of 1-PI Green functions $\Gamma$: 
\begin{equation}
\label{wgenchi}
W^{gen} \Gamma = \chi 
\partial_{\xi} \int \left\{ z^{-1} Y_1 \varphi_2 
                             -z Y_2 (\varphi_1 - \hat{\xi}_A
                             \frac{m}{e}) \right\}
\end{equation}
 In the following we
don't want to rely on special properties of a specific renormalization 
scheme, instead we will try to work scheme-independently (as far as possible).
Accordingly we only assume that in the scheme used the action principle
holds. This action principle tells us that
\begin{equation}
\label{gl4.13}
W ^{gen} \Gamma = \tilde{\Delta} \cdot \Gamma \; ,
\end{equation}
where $\tilde{\Delta}$ is a local integrated insertion with dimension less
than or equal to four, $\phi \pi$-charge $0$ and odd under charge
conjugation.
Because the Ward identity has already been established at $\chi = 0$
(\ref{wgen})  $\tilde \Delta $ has to depend explicitly
on $\chi $,
\begin{equation}
\label{qap}
\tilde \Delta = \chi {\tilde \Delta}_{-}  \; ,
\end{equation}
where ${\tilde \Delta} _{-} $ has  $\phi \pi$-charge $-1$. \\
The second ingredient, needed for the proof, is a certain transformation
behaviour of $W^{gen}$  (\ref{gl4.12})
with respect to BRS invariance: Acting with
$\hat W^{gen}$  (\ref{gl4.1}) on the ST identity (\ref{gl3.2}) yields 
\begin{equation}
\label{cons}
\hat W^{gen} {\cal S} (\Gamma ) =  s_{\Gamma} (\hat
W^{gen} \Gamma) - \chi( \partial
_ \xi \hat W^{gen} ) \Gamma = 0 \; .
\end{equation}
The operators (\ref{chiop}) are chosen in 
such a way as to satisfy the consistency
condition (\ref{cons}) by construction. Therefore one derives for the
$\chi $-enlarged operator $W^{gen}$ (\ref{gl4.12}) the consistency condition
\begin{eqnarray}
\label{gl4.14}
0 & = & W^{gen} {\cal S} (\Gamma) = 
s_{\Gamma} (W^{gen} \Gamma) \nonumber \\
& & \! \! \! \! - \chi \left[ \partial_{\xi} \int \left\{
  -z^{-1} \varphi_2 \frac{\delta}{\delta \varphi_1} +
   z (\varphi_1 - \hat{\xi}_A \frac{m}{e} )
     \frac{\delta}{\delta \varphi_2} -
   z Y_2 \frac{\delta}{\delta Y_1} +
   z^{-1} Y_1 \frac{\delta}{\delta Y_2} \right\} \right] \Gamma \; , 
\end{eqnarray}
where $s_{\Gamma}$ is given by:
\begin{equation}
\label{gl4.15}
s_{\Gamma} = \int \left\{ \partial_{\mu} c \frac{\delta}{\delta A_{\mu}}
  + B \frac{\delta}{\delta \bar{c}}
+ \frac{\delta \Gamma}{\delta \underline{Y}}
  \frac{\delta}{\delta \underline{\varphi}}
+ \frac{\delta \Gamma}{\delta \underline{\varphi}}
  \frac{\delta}{\delta \underline{Y}}
+ \underline{q} \frac{\delta}{\delta \underline{\hat{\varphi}}}
\right\} + \chi \partial_{\xi}
\end{equation}
Furthermore one calculates: 
\begin{eqnarray}
\label{sgdelta}
s_{\Gamma}\chi \Delta_{br} &\hspace{-3mm} = & \hspace{-3mm}
-\chi  s_{\Gamma} 
  \partial_{\xi} \int \left( z^{-1} Y_1 \varphi_2 
                             -z Y_2 (\varphi_1 - \hat{\xi}_A
                              \frac{m}{e}) \right) \nonumber\\
&\hspace{-3mm}= &\hspace{-3mm}
  - \chi \left[ \partial_{\xi} \int \! \! \left\{
  z^{-1} \varphi_2 \frac{\delta}{\delta \varphi_1} -
   z (\varphi_1 - \hat{\xi}_A \frac{m}{e} )
     \frac{\delta}{\delta \varphi_2} +
   z Y_2 \frac{\delta}{\delta Y_1} -
   z^{-1} Y_1 \frac{\delta}{\delta Y_2} \right\} \right] \Gamma  
\end{eqnarray}

Please note that the derivation of the consistency condition (\ref{gl4.14})
as well as 
(\ref{sgdelta}) only uses the invariance of the vertex functional $\Gamma$
with respect to the 
Slavnov-Taylor identity, and does not rely on any explicit
expressions for the Green functions.
Combining (\ref{gl4.14})    and (\ref{sgdelta}) one gets 
\begin{equation}
\label{conschi}
s_{\Gamma} ( W^{gen} \Gamma - \chi \Delta_{br} ) = 0
\end{equation}
which is valid to all orders of perturbation theory and constrains the
breaking of the Ward identity (\ref{gl4.13}), (\ref{qap}) 
to be $s_{\Gamma}$-invariant.
\begin{equation}
\label{conschi1}
s_{\Gamma} \bigl(\chi {\tilde \Delta}_-   \cdot \Gamma -\chi \Delta_{br}
\bigr)= 0
\end{equation}
 From here on we proceed by induction order by order in perturbation
theory in order to prove the $\chi$-enlarged Ward identity of rigid
symmetry. The Ward identity has been established in the tree aproximation
(\ref{wthooft}) and according to the quantum action principle
 the WI is broken at least by
a local polynomial in the fields in 1-loop order:
\begin{equation}
\label{gl4.15a}
\biggl( W^{gen} \Gamma \biggr)^{(\leq 1)}
- \chi \Delta_{br}^{(\leq 1)} = \chi \Delta _-^{(1)}
\end{equation}
The parameters $z, \xi_A, \hat \xi_A$ appearing in (\ref{gl4.15a}) are uniquely determined
at $\chi = 0$ and expanded including 1-loop order.
Applying the operator $s_{\Gamma} $ to equation (\ref{gl4.15a}) and
using the consistency condition (\ref{conschi1}) gives in
1-loop order
\begin{equation}
\label{cons1l}
s_{\Gamma} \chi \Delta _-^{(1)} = s _{\Gamma _{cl}} \chi \Delta_-^{(1)}
+ O(\hbar^2) = 0 \; .
\end{equation}
Solving (\ref{cons1l}) is a purely classical problem now. The list of
independent polynomials constituting $\Delta_ -^{(1)}$ 
can be given in the following form:
\begin{eqnarray}
\label{gl4.21}
{\Delta}_-^{(1)} 
& = & {\textstyle \int} \left\{ w_1 Y_1 \varphi_2 +
  w_2 Y_1 \hat{\varphi}_2 + w_3 Y_2 + w_4 Y_2 \varphi_1 +
  w_5 Y_2 \hat{\varphi}_1 \right. \nonumber \\
& & \left. + w_6 \bar{c} + w_7 (-x Y_2 + e\bar{c} (\bar{\varphi}_1 + v)) 
    + w_8 \bar{c} \hat{\varphi}_1
    + w_9 \bar{c} A^2 \right. \nonumber \\
& & \left. + w_{10} \bar{c} \varphi_1^2 
+ w_{11} \hat{\varphi}_1 (-x Y_2 + e\bar{c} (\bar{\varphi}_1 + v)) 
    + w_{12} \bar{c} \hat{\varphi}_1^2
+ w_{13} \bar{c} \varphi_2^2 \right. \nonumber \\
& & \left. + w_{14} \hat{\varphi}_2 (-x Y_1 - e\bar{c} \bar{\varphi}_2 ) 
+ w_{15} \bar{c} \hat{\varphi}_2^2 \right\}
\end{eqnarray}
All the coefficients $w_i$ are purely 1-loop.\\
Acting with $s_{\Gamma _{cl}}^{\chi = 0} $ on (\ref{gl4.21}) yields:
\begin{eqnarray}
\label{deltavar}
s_{\Gamma _{cl}}^{\chi = 0} {\Delta}_-^{(1)} & = & \int
\biggl( w_1 \bigl(\varphi_2 {\delta \Gamma_{cl} 
\over \delta \varphi_1  } 
- Y_1 {\delta \Gamma_{cl} \over \delta Y_2} \bigr) 
+ w_4 \bigl(\varphi_1 {\delta\Gamma_{cl}  \over \delta \varphi_2  } 
- Y_2 {\delta \Gamma_{cl} \over \delta Y_1 }\bigr) 
+ w_3 {\delta\Gamma_{cl}  \over \delta \varphi _2}  \\
& & 
+ w_{14} \bigl(\hat \varphi_2 {\delta \Gamma_{cl} 
\over \delta \hat \varphi_1  } 
+q_2 {\delta \Gamma_{cl} \over \delta q_1} \bigr) 
+ w_{11} \bigl(\hat \varphi_1 {\delta\Gamma_{cl}  \over \delta \hat
\varphi_2  } 
+q_1 {\delta \Gamma_{cl} \over \delta q_2 }\bigr) 
+ w_7 {\delta\Gamma_{cl}  \over \delta \hat \varphi _2} \nonumber \\
&&\,+  w_ 2\bigl(\hat \varphi_2 {\delta \Gamma_{cl} 
\over \delta \varphi_1  } 
- Y_1 q_2 \bigr) 
+ w_5 \bigl(\hat \varphi_1 {\delta\Gamma_{cl}  \over \delta \varphi_2  } 
- Y_2 q_1 \bigr) \nonumber \\
& & \,+ w_6 B + w_8 (B \hat \varphi _1 - \bar c q_1 ) + 
w_{12}(B \hat \varphi _1^2
-\bar c  s \hat \varphi _1^2 ) +
w_{15}(B \hat \varphi _2^2
-\bar c  s \hat \varphi _2^2 )  \nonumber \\
&& \,+ w_9 (B A^2 - \bar c sA^2 ) + 
w_{10}(B  \varphi _1^2
-\bar c  s  \varphi _1^2 ) +
w_{13}(B  \varphi _2^2
-\bar c  s  \varphi _2^2 ) \biggl)  \nonumber 
\end{eqnarray}
 The polynomials in (\ref{deltavar}) are not all independent as it is seen
from the classical Ward identity
\begin{equation}
 W \Gamma _{cl} \Big|_{\chi =0} = 0 \; .
\end{equation}
This means that there is a $s_\Gamma $-invariant in the basis of 
the $\Delta^{(1)} _- $, which could give rise to a $\chi $-anomaly 
of the global Ward identity. 
 The operators of the first two lines just constitute $\hat W $
and therefore one has to eliminate one of these polynomials via the
classical Ward identity, for instance $\varphi_2 {\delta \Gamma_{cl} 
\over \delta \varphi_1  } 
- Y_1 {\delta \Gamma_{cl} \over \delta Y_2} $.
The remaining 14 polynomials are independent and according to the 
consistency condition (\ref{cons1l}) their coefficients have to
vanish. Therefore the breaking of the Ward identity is restricted 
to one $s_\Gamma $-invariant whose coefficient is not available by
algebraic consistency:
\begin{equation}
\label{anomaly}
(W^{gen}\Gamma ) ^{(\leq 1)} = \chi \Delta _{br}^{(\leq 1)}  + w_1 \chi \int
\Bigl((
Y_1 -e \bar c \hat \varphi_2)\bar \varphi_2 - \bigl(
Y_2 + e \bar c( \hat \varphi_1 -
\frac me \xi_A)\bigr)(\bar \varphi _1 + \frac me) \Bigr)
\end{equation}
The coefficient $w_1$ has to be determined by an explicit test. Testing eq.\
(\ref{anomaly}) with respect to
 $Y_1 \varphi_2$ and $Y_2 \varphi _1$  at an asymptotic momentum $p^2_\infty
\gg m^2_i $, where the three point
functions disappear, the coefficient $ w_1$ is easily determined,
\begin{eqnarray}
\label{gl4.30}
\Gamma^{(1)}_{Y_1 \varphi_1}(p^2_\infty) + \Gamma^{(1)}_{Y_2 \varphi_2}(p^2
_\infty) & = &
 - \chi  w_1  \; , \\
\Gamma^{(1)}_{Y_1 \varphi_1} (p^2_\infty)+ \Gamma^{(1)}_{Y_2 \varphi_2}(p^2
_\infty) & = &
  \chi w_1 \; , \nonumber
\end{eqnarray}
and therefore
\begin{equation}
\label{gl4.31}
w_1 = 0 \; .
\end{equation}
The equations (\ref{anomaly})   and (\ref{gl4.31}) establish the $\chi $-dependent
Ward identity of rigid symmetry at 1-loop order as suggested by the
classical approximation.

It is clear how the induction proof has to be finished: We assume
 the validity of the
WI (\ref{wgenchi}) in order $n$ of $\hbar$.
We conclude, passing through exactly the same steps as above, that the 
WI also holds at order $n+1$. Hence we have proven to all orders:
\begin{equation}
\label{gl4.41}
W^{gen} \Gamma = \chi \Delta_{br}
\end{equation}

\newsection{The local Ward identity}
We conclude the general treatment of the $\chi$-dependence in the Abelian
Higgs model by constructing a local $\chi$-dependent WI which expresses
the invariance of Green functions under deformed local gauge transformations
and simultaneously governs the $\xi$-dependence of these Green functions.
In ref.\ \cite{KS} it was shown that at $\chi =0$ the 1-PI Green functions 
satisfy the local Ward identity:
\begin{equation}
\label{wardloc}
\bigl( {(e+ \delta e)}  w^{gen} (x) -
  \partial_{\mu} \frac{\delta}{\delta A_{\mu}} \bigr) \Gamma\Big|_{\chi= 0} =
\Box B 
\end{equation}
The local ($\chi$-dependent) operator $w^{gen} $ is defined 
 from the rigid one
(\ref{gl4.12}) by taking away the integration:
\begin{equation}
\label{gl5.1}
W^{gen} = \int d^4x \; w^{gen} (x)
\end{equation}
$\delta e $ is of order $\hbar$ and is determined by 
the normalization condition imposed for fixing the coupling $e$ in
higher order perturbative calculations.

The main result in deriving the local $\chi$--dependent Ward identity
is the proof, that the overall normalization factor
of the matter transformations $e+\delta e $ has to be independent
of the gauge parameter to all orders of perturbation theory.
This expresses the constraint we have found for the 
$\xi $--dependence of the vertex function
 $\Gamma _{A_\mu \varphi_1 \varphi_2} $ from the ST identity (\ref{vertex})
at the level of the Ward identity, where it is easily manageable in 
concrete calculations.

The derivation of the local $\chi$-dependent  Ward identity relies on
the same two ingredients as the one of the rigid Ward identity,
 namely the action principle
and the transformation behaviour
 of the local Ward operator $w^{gen}(x)$ (\ref{gl5.1}) under BRS
transformations. 
Combining the information gained about the transformation of the vertex 
functional under rigid symmetry (\ref{wgenchi}) and about the existence of
the
local Ward identity at $\chi = 0$ (\ref{wardloc})  the action  principle
tells us that at $\chi \neq  0 $ the local Ward identity is at least broken
by the local polynomial $\chi D_{br} (x) $ and the divergence of
a $\chi $-dependent current $\chi \partial _\mu \jmath ^\mu$, i.e.:
\begin{equation}
\label{wbreak}
\bigl( {(e+ \delta e)}  w^{gen} (x) -
  \partial_{\mu} \frac{\delta}{\delta A_{\mu}} \bigr) \Gamma = 
\Box B + (e + \delta e) \chi D_{br} (x) + \chi [\partial _\mu \jmath ^\mu]
\cdot \Gamma
\end{equation}  
 $D_{br}(x) $
 is defined to be the non-integrated breaking term $\Delta_{br}$
(\ref{gl4.11}) of the rigid WI:
\begin{equation}
\label{gl5.3}
D_{br}(x)= \partial_{\xi}  \left( z^{-1} Y_1 \varphi_2 
                             -z Y_2 (\varphi_1 - \hat{\xi}_A
                             \frac{m}{e}) \right) \; \; , \; \;
\Delta_{br} = \int d^4x \; D_{br} (x)
\end{equation}
The current $\jmath _{\mu}$ has 
$\phi\pi$-charge -1, dimension 3 and is odd under
charge conjugation. Inspecting the quantum numbers of the fields in question
it turns out
that there is only one field polynomial satisfying the above requirements,
\begin{equation}
\jmath _{\mu} = u \partial _{\mu} \bar c \; ,
\end{equation}
which is linear in the propagating fields and therefore a trival insertion to all orders.

Turning now to the consistency conditions for the local Ward identity
one derives  in direct analogy to (\ref{gl4.14})--(\ref{conschi}):  
\begin{eqnarray}
\label{gl5.5}
& 0 =   w^{gen} (x) {\cal S} (\Gamma) =
  s_{\Gamma} ( w^{gen} (x) \Gamma - \chi D _{br}(x))  \; \; , &
 \\
& 0 = \partial _\mu {\delta \over \delta A_\mu } {\cal S} (\Gamma )
= s_{\Gamma} \Bigl( \partial _\mu {\delta \over \delta A_\mu } \Gamma \Bigr)
&
\end{eqnarray}
Applying
  the $\chi$-dependent $s_\Gamma$-operator (\ref{gl4.15})
to eq.\ (\ref{wbreak}) and using the consistency condition (\ref{gl5.5})
one finds the following algebraic constraint valid to all orders of
perturbation theory:
\begin{equation}
\label{consloc}
\chi\Bigl( \partial _\xi (e +\delta e) \Bigr) \, w^{gen} \Gamma =
-\chi s_\Gamma u \Box \bar c =  - \chi u \Box B
\end{equation}
In the tree approximation, where $\partial _\xi e =0 $ is trivially fulfilled,
it follows
\begin{equation}
u^{(0)}= 0\; ,
\end{equation}
which is in agreement with the explicit determination.

For higher orders we have to note that the two insertions
$w^{gen} \Gamma $ and $\Box B $ are independent and have to
vanish separately in order to fulfill the consistency 
condition (\ref{consloc}). This implies to all orders
\begin{equation}
u = 0 \qquad \hbox{and} \qquad \partial _\xi \delta e = 0 \; \; .
\end{equation}
The independence of the two insertions is seen immediately by calculating
the lowest order of $w^{gen} \Gamma $ at $\chi = 0$:
\begin{equation} 
w^{gen} \Gamma \Big|_{\chi = 0} = \partial ^\mu [j_{\mu}^{matter}]\cdot \Gamma
\end{equation}
with
\begin{equation}
\label{gl5.9}
j_{\mu}^{matter} = \bar{\varphi}_2 \partial_{\mu} \bar{\varphi}_1
  - (\bar{\varphi}_1 + \hbox{$\frac me$}) \partial_{\mu} \bar{\varphi}_2 
  + e A_{\mu} (\bar{\varphi}_2^2 + (\bar{\varphi}_1 + 
\hbox{$\frac me$})^2) + O(\hbar)
\end{equation}    
$j_{\mu}^{matter}$ is -- in contrast to $\Box B$ -- non-trivially inserted    into higher
orders' loop diagrams.

Therefore the local Ward-identity
\begin{equation}
\label{result}
\left( {(e+ \delta e)}  w^{gen} (x) -
  \partial_{\mu} \frac{\delta}{\delta A_{\mu}} \right) \Gamma = 
\Box B + (e + \delta e) \chi D_{br} (x) 
\end{equation}
with the restriction
\begin{equation}
\partial_\xi(e+ \delta e )   = 0
\end{equation}
is established to all orders of perturbation theory.
Because of the spontaneous symmetry breaking, in $w^{gen}$ there appear
the derivatives with respect to $\varphi_2 $ and $ \hat \varphi  _2 $.
Hence local gauge invariance is non-trivially 
formulated at the level of Green
functions when compared to classical gauge invariance acting on fields.

As we have already mentioned the main result is the $\xi$-independence
of the factor $e+ \delta e$ appearing in the local Ward identity.
This is a highly non-trivial result in
higher orders of perturbation theory and can be deduced only with the
formalism of a BRS-varying gauge parameter $\xi $.  Finding in a
explicit calculation $\delta e $ to be $\xi$-dependent means that the
normalization conditions imposed are not in agreement with the $\chi
$-enlarged Slavnov-Taylor identity and it is suggested that under such
circumstances the $\xi $-independence of the S-matrix cannot be proven
which in turn is a desaster for the definition of the theory.

Therefore it is obvious that we can also use the local Ward identity
in order to fix the coupling $e$, i.e.\ we require the local Ward identity
to be exact to all orders of perturbation theory:
\begin{equation}
\label{normeward}
\bigl(({e}  w^{gen} (x) -
  \partial_{\mu} \frac{\delta}{\delta A_{\mu}} ) \Gamma \Big|_{\chi
=0}
= \Box B 
\end{equation}
This last eq. reads for the vertex function in question:
\begin{eqnarray}
\label{evertex}
& e
\Bigl( z^{-1} \Gamma _{\varphi_1 \varphi_1} (p_1^2) -
         z   \bigl(   \Gamma _{\varphi_2 \varphi_2} (p_2^2) 
       -  \hat \xi_A \hbox{$\frac me $ }
 \Gamma _{\varphi_2 \varphi_1 \varphi_2}
 (p, p_1 ,p_2 )
       -   \xi_A \hbox{$\frac me $ }
 \Gamma _{\hat \varphi_2 \varphi_1 \varphi_2}
             (p, p_1 ,p_2 ) \bigr) \Bigr) & \nonumber \\
& = i(p_1^\mu + p_2^\mu ) \Gamma _{A^\mu \varphi_1
        \varphi_2 }(p, p_1 , p_2 ) & \\
& \hbox{with}
\quad p^\mu+p_1^\mu+p_2^\mu = 0 & \nonumber
\end{eqnarray} 
As long as we do not apply a normalization condition for fixing the
coupling $e$, the vertex $
  \Gamma _{A^\mu \varphi_1
        \varphi_2 }(p, p_1 , p_2 ) $
is only  determined up to a
local contribution, i.e. a finite counterterm, which can be added in each
 order:
\begin{equation}
{\Gamma '} _{A^\mu \varphi_1
        \varphi_2 }
^{(n)}(p, p_1 , p_2 ) = \Gamma^{(n)} _{A^\mu \varphi_1
        \varphi_2 }(p, p_1 , p_2 ) - i e (p_1^\mu - p_2 ^\mu ) \delta z_e
 ^{(n)}(\xi)
\end{equation}
 ${\Gamma '} _{A^\mu \varphi_1
        \varphi_2 }   $ and
 ${\Gamma } _{A^\mu \varphi_1
        \varphi_2 } $
are equivalent solutions of the 
 ordinary ST-identity at $\chi = 0$ for an arbitrary $\delta z_e (\xi)$.

 It was shown, however, that the dependence on
the gauge parameter cannot be arbitrarily adjusted. Instead, this has to be done in such a way
that the ST identity holds at $\chi \neq 0  $. Equivalently
-- and this is the remarkable point of the analysis --  one can require
that the $\xi $-independence of the factor $e+ \delta e $ is not ruined.
This is trivially fulfilled if we require $\delta e =0$.
It is seen that the local Ward identity (\ref{normeward}) uniquely determines
$\delta z_e $, if all the other normalization conditions are applied, rigid 
invariance is established and the Slavnov-Taylor identity at $\chi=0 $
holds.

Hence taking the on-shell conditions together with the 
requirement ``Ward identity exact to all orders'' (\ref{normeward})
the determination of the gauge parameter  dependence is
in agreement with the Slavnov-Taylor identity at $\chi \neq 0$.
In explicit calculations these conditions ensure that gauge parameter
dependence is handled correctly.

Due to  the spontaneous symmetry breaking the requirement ``Ward identity
of local invariance being exact'' is not immediately related  to 
a physical interpretation of the coupling, as can be done in QED.
There the same requirement is even physical, because the
coupling is  seen to be the fine structure constant in the Thompson 
limes. Using the Ward identity of rigid symmetry  and the
complete on-shell conditions for the Higgs particle 
(\ref{massnorm}), (\ref{wavenorm})
one finds:
\begin{equation}
\left.
\frac {p^\mu \Gamma_{A^\mu \varphi_1 \varphi_2}(p,p_1,p_2)}{p_1^2 - p_2^2}
\right|_{p^2 = 0 \atop p_1^2 =p_2^2 = m_H^2} = e z
\end{equation}
with $z$ being determined from the rigid Ward identity.
A similar expression can be derived for the on-shell conditions, when
the residuum of the Higgs is
 fixed at an arbitrary normalization point $\kappa$.
In any case the parameter $e$ has to be adjusted to its physical value
by calculating  a physical process, which is manifestly gauge parameter
independent by construction. 

\newsection{Conclusions}
In the present paper we have investigated the renormalization of the
Abelian Higgs model including a BRS-varying gauge parameter.  
The techniques we have presented relate gauge parameter dependence of
the Green functions to the evaluation of a class of ghost diagrams via
the enlarged Slavnov-Taylor identity. This procedure is well-defined
and can be carried out by just implementing additional Grassmann valued
vertices into the action. The diagrams, which determine gauge
parameter dependence, have a much simpler structure
when compared to the original
ones, in general they are also less divergent.

The importance of the algebraic method for controlling gauge parameter
dependence is founded in the fact that the Green functions
constructed with a BRS-varying gauge parameter are determined in a
way which ensures all the physical quantities to be gauge parameter
independent just by construction. Especially it is seen, that
the normalization conditions, which fix the physical parameters of the
model (like the masses of the vector and the Higgs and the coupling
constant), cannot be chosen arbitrarily concerning gauge parameter
dependence. The advantage of such a construction is obvious, because
one is forced to adjust the counterterms correctly already in the
procedure of renormalizing the 1-PI Green functions. Otherwise, if one
does not introduce a BRS-varying gauge parameter, it can happen, that
gauge parameter dependence enters the S-matrix by a wrong adjustment
of counterterms. Then at the end, this dependence
has to be removed order by order in
perturbation theory by carefully adjusting the gauge parameter
dependent part of the counterterms once again. Such a treatment is not
transparent and difficult to control, especially
if the model contains several couplings and parameters.

In the  Abelian Higgs model, which we have chosen
as the simplest example of a  spontaneously broken gauge theory, the
results of the algebraic method are quite impressive:
Inspection of the Slavnov-Taylor identities and the corresponding
diagrams tells us that the transversal part of the vector self energy is
independent of the gauge parameter to all orders of perturbation
theory.  Furthermore, it is seen that the on-shell conditions for the
physical particles are in complete agreement with the restrictions 
arising from the enlarged Slavnov-Taylor identity. In the Abelian
Higgs model one is able to derive a local Ward identity of gauge
symmetry. We have proven, that the construction prohibits the gauge
parameter to enter the overall normalization factor of 
the matter transformations (\ref{result}), which in lowest 
order coincides  with the coupling. Vice versa, making the
Ward identity exact to all orders, ensures by itself that gauge
parameter dependence is treated correctly.

The techniques applied here are universal and can be generalized to
more complex models and situations, whenever one is interested in an
explicit knowledge about gauge parameter dependence or 
whenever one wants to
construct and to analyze gauge parameter independent quantities.
The S-matrix is the most important object for such considerations.
In this context, the analysis of
gauge parameter dependence
in the case when the thoery contains unstable particles
is outstanding: The Slavnov-Taylor identity holds as it is and it
remains to study the consequences thereof
in order to arrive at a proper definition of the
mass and width of the unstable particles.
\\[1cm]

{\it Acknowledgements }\hspace*{0.5cm}  We are 
grateful to K.~Sibold for numerous helpful
comments and a critical reading of the mansucript. One of us (E.K.) wants to
thank G.~Weiglein for useful discussions.\\[3cm]

\renewcommand{\theequation}{A.\arabic{equation}}
\setcounter{equation}{0}
\noindent
{\large \bf Appendix A}\\[0.5cm]
In this appendix we present (without any calculation) the general solution of
the gauge condition (\ref{gl3.3}) and the (usual, that is $\chi$-independent)
ST identity
\begin{equation}
\label{gla.1}
{\cal S} (\hat{\Gamma}) = \int \left\{
  \partial_{\mu} c \; \frac{\delta \hat{\Gamma}}{\delta A_{\mu}}
\; + \; B \; \frac{\delta \hat{\Gamma}}{\delta \bar{c}}
\; + \; \frac{\delta \hat{\Gamma}}{\delta \underline{Y}} \;
        \frac{\delta \hat{\Gamma}}{\delta \underline{\varphi}}
\; + \; \underline{q} \;
        \frac{\delta \hat{\Gamma}}{\delta \underline{\hat{\varphi}}}
\right\} = 0
\end{equation}
in the classical
 approximation. It turns out that the classical solution can be decomposed
as follows
\begin{equation}
\label{gla.2}
\hat{\Gamma}_{cl}^{gen} = \Lambda (A_{\mu}, \bar{\varphi}_1, \bar{\varphi}_2 )
\; + \; \Gamma_{g.f.} \; + \; \Gamma_{\phi \pi} \; + \; \Gamma_{e.f.} \; ,
\end{equation}
where
\begin{equation}
\label{gla.3}
\bar{\varphi}_i = \varphi_i - x_i \hat{\varphi}_i \; , \; i = 1,2 \; .
\end{equation}
The matter part
 $\Lambda = \Lambda (A_{\mu}, \bar{\varphi}_1, \bar{\varphi}_2 )$ 
is given by
\begin{eqnarray}
\label{gla.4}
\Lambda & = &
  \int \left\{ -\; \frac{z_A}{4} F_{\mu \nu} F^{\mu \nu} \; + \;
  \frac{1}{2} z_1 (\partial_{\mu} \bar{\varphi}_1 )
                  (\partial^{\mu} \bar{\varphi}_1 ) \; + \;
\frac{1}{2} z_2 (\partial_{\mu} \bar{\varphi}_2 )
                (\partial^{\mu} \bar{\varphi}_2 ) \right. \nonumber \\
& & \left. + \; z_e e \sqrt {z_1} \sqrt{z_2} \sqrt{z_A} \left( 
                (\partial_{\mu} \bar{\varphi}_1 ) \bar{\varphi}_2 \; - \;
                \bar{\varphi}_1 (\partial_{\mu} \bar{\varphi}_2 ) \right)
                A^{\mu} \; + \;
\frac{1}{2} z_e^2 {e}^2 z_A  (z_1 \bar{\varphi}_1^2 \; + \;
                               z_2 \bar{\varphi}_2^2) A_{\mu} A^{\mu}
\right. \nonumber \\
& & \left. + \; \frac{1}{2} z_m m^2 z_A  A_{\mu} A^{\mu} \; - \;
    \sqrt{ z_2}\sqrt{z_m} m \sqrt{z_A}
(\partial_{\mu} \bar{\varphi}_2 ) A^{\mu} \; + \;
z_e e \sqrt{ z_m} m \sqrt {z_1} z_A
\bar{\varphi}_1 A_{\mu} A^{\mu} \right. \nonumber \\
& & \left. + \; \frac{1}{2} \mu^2
    (z_1 \bar{\varphi}_1^2 + 2 \sqrt{z_1} 
\frac {\sqrt{z_m} m}{{z_e} e} \bar{\varphi}_1 +
     z_2 \bar{\varphi}_2^2) \right. \nonumber \\
& & \left. \; - 
\frac 18 \frac{z_{m_H}m_H^2}{z_m m^2}z_e^2 e^2 (z_1 \bar{\varphi}_1^2 + 
2 \sqrt{z_1} \frac {\sqrt{z_m} m}{{z_e} e} 
 \bar{\varphi}_1 +
z_2 \bar{\varphi}_2^2)^2 \right\}
\end{eqnarray}
The gauge fixing part $\Gamma_{g.f.}$ directly results from integrating
the gauge condition (\ref{gl3.3}):
\begin{equation}
\label{gla.5}
\Gamma_{g.f.} = \int \left\{ \frac{1}{2} \xi B^2 + 
  B \partial A - 
e B \left[ (\hat{\varphi}_1 - \xi_A \frac{m}{e} ) \varphi_2 -
            \hat{\varphi}_2 (\varphi_1 - \hat{\xi}_A \frac{m}{e}) \right]
\right\}
\end{equation}
The remaining two parts, the external field part $\Gamma_{e.f.}$ and the
$\phi \pi$-part $\Gamma_{\phi \pi}$, have the form
\begin{equation}
\label{gla.6}
\Gamma_{e.f.} = \int \{ Y_1 (-{e}z_e \sqrt {\frac{z_2}{z_1}} 
\sqrt {z_A} \bar{\varphi}_2 c +
                                  x_1 q_1) \; + \;
Y_2 ({e}z_e \sqrt {\frac{z_1}{z_2}} \sqrt{z_A} (\bar{\varphi}_1 + 
\frac {\sqrt{z_m} m}{\sqrt{z_1} {z_e} e} ) c + x_2 q_2) \}
\end{equation}
and
\begin{eqnarray}
\label{gla.7}
\Gamma_{\phi \pi} & = & \int \{
  - \bar{c} \Box c \; + \; e\bar{c} (q_1 \varphi_2 - q_2 (\varphi_1 -
                               \hat{\xi}_A \frac{m}{e})) \nonumber \\
& & + \; e \bar{c} (\hat{\varphi}_1 - \xi_A \frac{m}{e})
                   (z_e e \sqrt{\frac{z_1}{z_2}} \sqrt{z_A} 
 (\bar{\varphi}_1 + \frac{\sqrt{z_m} m}{\sqrt{z_1} z_e e} ) c + x_2 q_2 )
    \nonumber \\
& & - \; e \bar{c} \hat{\varphi}_2 
 (-z_e e \sqrt{\frac{z_2}{z_1}} \sqrt{z_A} \bar{\varphi}_2 c
                                           + x_1 q_1) \} \; .
\end{eqnarray}
This general solution of the ST identity (\ref{gla.1}) contains quite a number
of so far free parameters, namely the wave function normalizations $z_1,
z_2$ and $z_A$, the mass renormalizations of the vector and the Higgs-particle, i.e.
$z_m, z_{m_H}$, the coupling renormalization $z_e$, the parameters $x_1, x_2$,
the t'Hooft gauge parameter $\xi_A$ and the parameter $\hat{\xi}_A$, which
are not prescribed by the ST identity (\ref{gla.1}) and which therefore have
to be fixed by appropriate
normalization conditions to all orders (see section 2).\\[2cm]

\renewcommand{\theequation}{B.\arabic{equation}}
\setcounter{equation}{0}
\noindent
{\large \bf Appendix B}\\[0.5cm]
According to the considerations of section 5 (see especially 
(\ref{higgs1l})) 
 gauge parameter dependence of the
Higgs self-energy is completely governed by the non-local
contributions to the vertex $\partial_{\chi} \Gamma_{Y_1 \varphi_1}$.
In the following we sketch the respective one-loop diagrams:

\unitlength1mm
\begin{picture}(170,170)
\put(-5,-43){\psfig{file=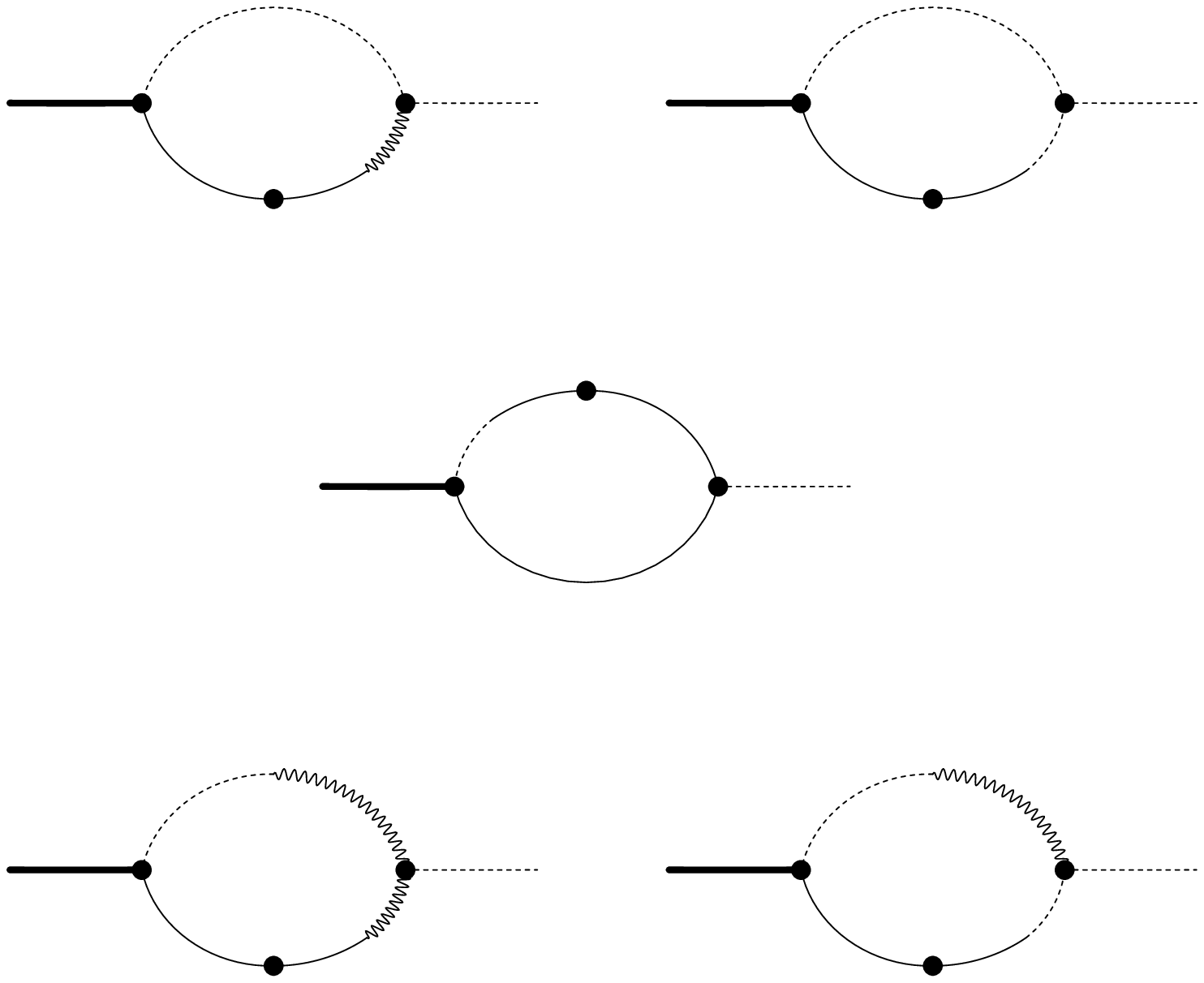,width=170mm}}

\put(10,146){$Y_1$}
\put(61,146){$\varphi_1$}
\put(35,158){$\varphi_2$}
\put(36,127){$\chi$}
\put(21,133){$c$}
\put(51,133){$A_{\mu}$}
\put(28,128){$\bar{c}$}
\put(44,128){$B$}

\put(95,146){$Y_1$}
\put(146,146){$\varphi_1$}
\put(120,158){$\varphi_2$}
\put(121,127){$\chi$}
\put(106,133){$c$}
\put(136,133){$\varphi_2$}
\put(113,128){$\bar{c}$}
\put(129,128){$B$}

\put(51,97){$Y_1$}
\put(102,97){$\varphi_1$}
\put(61,104){$\varphi_2$}
\put(69,109){$B$}
\put(77,110){$\chi$}
\put(85,109){$\bar{c}$}
\put(93,103){$c$}
\put(67,80){$c$}
\put(87,80){$\bar{c}$}

\put(10,47){$Y_1$}
\put(21,34){$c$}
\put(28,29){$\bar{c}$}
\put(36,28){$\chi$}
\put(44,29){$B$}
\put(51,34){$A_{\nu}$}
\put(61,47){$\varphi_1$}
\put(26,58){$\varphi_2$}
\put(46,57){$A_{\mu}$}

\put(95,47){$Y_1$}
\put(106,34){$c$}
\put(113,29){$\bar{c}$}
\put(121,28){$\chi$}
\put(129,29){$B$}
\put(136,34){$\varphi_2$}
\put(146,47){$\varphi_1$}
\put(111,58){$\varphi_2$}
\put(131,57){$A_{\mu}$}

\put(38,13){Figure 1: 1-loop order contributions to $\partial_{\chi}
\Gamma_{Y_1 \varphi_1}$}

\end{picture}

Please note that in the non-local contributions to the vertex
$\partial_{\chi} \Gamma_{Y_1 \varphi_1}$ there appear the mixed
propagators
\begin{eqnarray}
\label{mixpro}
G_{B A_{\mu}} (p,-p) & = & \frac{-p^{\mu}}{p^2 - \xi_A m^2} \; , \nonumber \\
G_{B \varphi_2} (p,-p) & = & \frac{-im}{p^2 - \xi_A m^2} \; , \\
G_{\varphi_2 A_{\mu}} (p,-p) & = & 
    \frac{-m (\xi - \xi_A) p^{\mu}}{(p^2 - \xi_A m^2)^2} \; . \nonumber
\end{eqnarray}

In the 't Hooft gauge $\xi_A = \xi$ the 
last two diagrams of figure 1 are absent (because
the $\varphi_2$-$A_{\mu}$-propagator vanishes), instead there is the
additional vertex $\chi \bar{c} \varphi_2 m$ and therefore:
\begin{picture}(60,170)
\put(-5,-43){\psfig{file=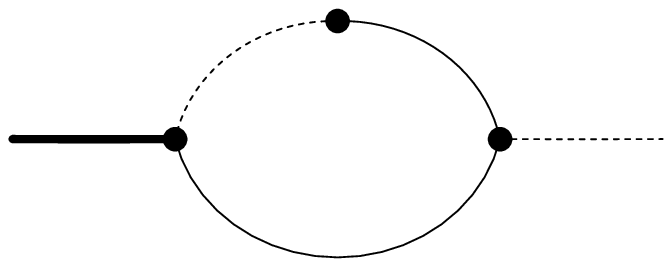,width=170mm}}

\put(51,146){$Y_1$}
\put(66,158){$\varphi_2$}
\put(75,159){$\chi m$}
\put(85,157){$\bar{c}$}
\put(93,152){$c$}
\put(102,146){$\varphi_1$}
\put(67,129){$c$}
\put(87,129){$\bar{c}$}

\put(29,114){Figure 2: Additional contribution in the 't Hooft gauge}

\end{picture}
\newpage
{

}
%
\end{document}